\documentclass[12pt,a4paper]{article}
\usepackage{graphicx}
\usepackage[T1]{fontenc}
\usepackage[utf8]{inputenc}
\usepackage{textcomp}
\usepackage[sc,osf]{mathpazo}
\usepackage{a4wide}  
\usepackage{latexsym,amsthm,amsfonts,amsmath,mathrsfs,amssymb}
\usepackage{dsfont}
\usepackage{accents}
\usepackage[nosort]{cite}
\usepackage{booktabs} % for much better looking tables
\usepackage[unicode,implicit]{hyperref}
\hypersetup{%
  pdftitle    = {T~duality and Wald entropy formula in the Heterotic Superstring
    effective action at first order in alpha'}
  pdfkeywords = {duality, T duality, strings, heterotic superstrings, entropy,
  black holes, Wald, compactification}
  pdfauthor   = {Zachary Elgood and Tomas Ortin},
  plainpages  = true,
  colorlinks  = true,
  citecolor   = blue,
  urlcolor    = red,
  linkcolor   = black
}
\newcommand{\hepth}[1]{{\tt
\href{http://www.arXiv.org/abs/hep-th/#1}{hep-th/#1}}}
\newcommand{\grqc}[1]{{\tt
\href{http://www.arXiv.org/abs/gr-qc/#1}{gr-qc/#1}}}

\newcommand{\arxiv}[1]{{\tt arXiv:\href{http://www.arXiv.org/abs/#1}{#1}}}
\newcommand{\doi}[1]{{\tt DOI:\href{http://dx.doi.org/#1}{#1}}}

\makeatletter
\@addtoreset{equation}{section}
\makeatother

\begin{document}

\begin{flushright}
\small
IFT-UAM/CSIC-20-067\\
\texttt{arXiv:2005.11272 [hep-th]}\\
May 23\textsuperscript{rd}, 2020\\
\normalsize
\end{flushright}

\vspace{1cm}

\begin{center}

{\Large {\bf T~duality and Wald entropy formula\\[.5cm] in the Heterotic Superstring
    effective action\\[.5cm] at first-order in $\alpha'$}}

\vspace{1.5cm}

\renewcommand{\thefootnote}{\alph{footnote}}

{\sl\large Zachary Elgood}\footnote{Email: {\tt zachary.elgood[at]uam.es}}
{\sl\large and Tom\'{a}s Ort\'{\i}n}\footnote{Email: {\tt tomas.ortin[at]csic.es}}

\setcounter{footnote}{0}
\renewcommand{\thefootnote}{\arabic{footnote}}

\vspace{1cm}

{\it Instituto de F\'{\i}sica Te\'orica UAM/CSIC\\
C/ Nicol\'as Cabrera, 13--15,  C.U.~Cantoblanco, E-28049 Madrid, Spain}

\vspace{1cm}

%%%%%%%%%%%%%%%%%%%%%%%%%%%%%%%%%%%%%%%%%%%%%%%%%%%%%%%%%%%%%%%%%%%%%%

{\bf Abstract}
\end{center}
\begin{quotation}
  {\small We consider the compactification on a circle of the Heterotic
    Superstring effective action to first order in the Regge slope parameter
    $\alpha'$ and re-derive the $\alpha'$-corrected Buscher rules first found
    in Ref.~\cite{Bergshoeff:1995cg}, proving the T~duality invariance of the
    dimensionally-reduced action to that order in $\alpha'$. We use Iyer and
    Wald's prescription to derive an entropy formula that can be applied to
    black-hole solutions which can be obtained by a single non-trivial
    compactification on a circle and discuss its invariance under the
    $\alpha'$-corrected T~duality transformations. This formula has been
    successfully applied to $\alpha'$-corrected 4-dimensional non-extremal
    Reissner-Nordstr\"om black holes in Ref.~\cite{Cano:2019ycn} and we apply it
    here to a heterotic version of the Strominger-Vafa 5-dimensional extremal
    black hole.}
\end{quotation}

\newpage
%%%%%%%%%%%%%%%%%%%%%%%%%%%%%%%%%%%%%%%%%%%%%%%%%%%%%%%%%%%%%%%%%%%%%%
%%%%%%%%%%%%%%%%%%%%%%%%%%%%%%%%%%%%%%%%%%%%%%%%%%%%%%%%%%%%%%%%%%%%%%
%%%%%%%%%%%%%%%%%%%%%%%%%%%%%%%%%%%%%%%%%%%%%%%%%%%%%%%%%%%%%%%%%%%%%%
%%%%%%%%%%%%%%%%%%%%%%%%%%%%%%%%%%%%%%%%%%%%%%%%%%%%%%%%%%%%%%%%%%%%%%
\pagestyle{plain}
%%%%%%%%%%%%%%%%%%%%%%%%%%%%%%%%%%%%%%%%%%%%%%%%%%%%%%%%%%%%%%%%%%%%%%
%%%%%%%%%%%%%%%%%%%%%%%%%%%%%%%%%%%%%%%%%%%%%%%%%%%%%%%%%%%%%%%%%%%%%%
%%%%%%%%%%%%%%%%%%%%%%%%%%%%%%%%%%%%%%%%%%%%%%%%%%%%%%%%%%%%%%%%%%%%%%
%%%%%%%%%%%%%%%%%%%%%%%%%%%%%%%%%%%%%%%%%%%%%%%%%%%%%%%%%%%%%%%%%%%%%%

\tableofcontents

%\newpage

%%%%%%%%%%%%%%%%%%%%%%%%%%%%%%%%%%%%%%%%%%%%%%%%%%%%%%%%%%%%%%%%%%%%%%
%%%%%%%%%%%%%%%%%%%%%%%%%%%%%%%%%%%%%%%%%%%%%%%%%%%%%%%%%%%%%%%%%%%%%%
%%%%%%%%%%%%%%%%%%%%%%%%%%%%%%%%%%%%%%%%%%%%%%%%%%%%%%%%%%%%%%%%%%%%%%
%%%%%%%%%%%%%%%%%%%%%%%%%%%%%%%%%%%%%%%%%%%%%%%%%%%%%%%%%%%%%%%%%%%%%%
\section{Introduction}
%%%%%%%%%%%%%%%%%%%%%%%%%%%%%%%%%%%%%%%%%%%%%%%%%%%%%%%%%%%%%%%%%%%%%%
%%%%%%%%%%%%%%%%%%%%%%%%%%%%%%%%%%%%%%%%%%%%%%%%%%%%%%%%%%%%%%%%%%%%%%
%%%%%%%%%%%%%%%%%%%%%%%%%%%%%%%%%%%%%%%%%%%%%%%%%%%%%%%%%%%%%%%%%%%%%%
%%%%%%%%%%%%%%%%%%%%%%%%%%%%%%%%%%%%%%%%%%%%%%%%%%%%%%%%%%%%%%%%%%%%%%

Superstring Theory is expected to be a consistent theory of Quantum
Gravity. Therefore, one would like to use it to study gravitational systems in
which quantum-mechanical effects are believed to play an important role, such
as black holes. In particular, one of the results that we expect from
Superstring Theory is a microscopical accounting of the entropy attributed to
them by macroscopic (thermodynamic) laws and calculations.

Achieving this result demands, first of all, black-hole solutions of
Superstring Theory whose macroscopic entropy can be computed. These are
classical solutions of the Superstring effective action. Then, if one manages
to associate the black-hole solution to a good Superstring Theory background
on which the theory can be quantized, the microscopic entropy can be
associated to the density of string states in that background.

In a seminal paper, \cite{Strominger:1996sh} Strominger and Vafa completed the
above program for a extremal, static, 3-charge 5-dimensional black-hole
solution of the type~IIB Superstring Theory at lowest order in the Regge slope
parameter $\alpha'$, identifying the associated type~IIB string background as
one with intersecting D1- and D5-branes with momentum flowing along the
intersection.  Strominger and Vafa argued that, although the black hole only
solved the zeroth-order in $\alpha'$ equations of motion, the higher-order
corrections could be made small enough by imposing conditions on the charges
carried by the black hole. Under those conditions, the microscopic and
macroscopic entropies (the later given simply by the one fourth of the area of
the event horizon) matched to lowest order in $\alpha'$.

Since $\alpha'$ is the square of the string length, the higher-order in
$\alpha'$ corrections to the string effective action, its solutions and the
properties of the solutions describe characteristic ``stringy'' deviations and
this makes their study most interesting. This study requires:

\begin{enumerate}
\item The knowledge of the higher-order terms in the string effective
  field-theory actions.
\item The construction of solutions of those effective actions with
  higher-order terms. These solutions can often be viewed as
  $\alpha'$-corrected zeroth-order solutions (recovered by setting
  $\alpha'=0$).
\item The computation of the physical properties of the $\alpha'$-corrected
  solutions.
\end{enumerate}

Terms of higher-order in $\alpha'$ are terms of higher order in curvatures and
their complexity grows rapidly with the power of $\alpha'$. This makes them
very difficult to compute and, consequently, our knowledge of the
$\alpha'$ corrections to the effective field theory actions of different
Superstring Theories is very limited. The $\alpha'$ corrections to the
Heterotic Superstring effective action are probably the best known, and they
have only been computed to cubic order (quartic in curvatures) in
Ref.~\cite{Bergshoeff:1989de}, using supersymmetry completion of the Lorentz
Chern-Simons terms \cite{Bergshoeff:1988nn}.\footnote{The equivalence of this
  effective action with previous results obtained in
  Refs.~\cite{Callan:1985ia,Gross:1986mw,Metsaev:1987zx,Hull:1987pc} was
  established in Ref.~\cite{Chemissany:2007he}.}

We can use, then, the Heterotic Superstring effective action given in
Ref.~\cite{Bergshoeff:1989de} for the next step: computing $\alpha'$
corrections to black-hole solutions.  As a matter of fact, the black-hole
solution studied by Strominger and Vafa in Ref.~\cite{Strominger:1996sh} can
also be considered as a zeroth-order solution of the Heterotic Superstring
effective action and it would certainly be interesting to compute its
$\alpha'$ corrections, at least to first order. Finding these corrections,
though, is a complicated problem. One of the problems is that the complete
Heterotic Superstring effective action with higher-order corrections has not
been compactified down to the 5 dimensions in which the black hole
lives.\footnote{A toroidal compactification to first order in $\alpha'$ but
  with no Yang-Mills fields has been recently constructed in
  Ref.~\cite{Eloy:2020dko}. The toroidal compactification with only Abelian
  Yang-Mills fields (which occur at first order in $\alpha'$) and no terms
  involving the torsionful spin connection (so the 10-dimensional action is
  that of $\mathcal{N}=1,d=10$ supergravity coupled to Abelian vector
  supermultiplets) was carried out in \cite{Maharana:1992my}. An earlier
  compactification of the Heterotic Superstring effective action to just $d=4$
  at zeroth-order in $\alpha'$ (so the 10-dimensional action is that of pure
  $\mathcal{N}=1,d=10$ supergravity) was carried out in
  \cite{Chamseddine:1980cp}.} Effective actions which would capture what are
believed to be the most relevant $\alpha'$ corrections in lower dimensions
have been proposed and used to compute corrections to black-hole solutions
(see, \textit{e.g.}~Ref.~\cite{Mohaupt:2000mj} and references
therein). Alternatively, in order to simplify the problem, it has been
proposed to work only with the near-horizon solution (see
\textit{e.g.}~Refs.~\cite{Sen:2007qy,Prester:2010cw} and references therein
and more recent work in the Type~IIA compactified on K3 setup
\cite{Pang:2019qwq,Chow:2019win}). It is fair to say that each of these
simplified approaches has problems of its own and that they do not offer a
complete picture of what the $\alpha'$-corrected black-hole solutions are
like.

Recently, a different approach for computing $\alpha'$ corrections without
making assumptions about the lower-dimensional effective actions or
considering only near-horizon limits has been proposed in
Ref.~\cite{Cano:2018qev}: since the 10-dimensional first-order in $\alpha'$
Heterotic Superstring effective action is known without any ambiguities (beyond
possible field redefinitions), first-order in $\alpha'$ corrections to
solutions should be directly computed in 10 dimensions using the uplift of 4-
or 5-dimensional solutions. Then, the $\alpha'$-corrected solutions can be
compactified back to 4- or 5-dimensions. This approach has been successfully
used to compute the first-order in $\alpha'$ corrections to 5- and
4-dimensional extremal black holes in Refs.~\cite{Cano:2018qev} and
\cite{Chimento:2018kop,Cano:2018brq,Cano:2019oma}, respectively and, more
recently, to 4-dimensional non-extremal Reissner-Nordstr\"om black holes in
Ref.~\cite{Cano:2019ycn}. The question of the regularity of the so-called
\textit{small black holes} has also been reviewed in
Ref.~\cite{Cano:2018hut,Ruiperez:2020qda} in the light of those results.

Having the $\alpha'$-corrected solutions we can compute their physical
properties. For black holes, these are their conserved charges and their
thermodynamical properties: entropy and temperature. The Hawking temperature is
always determined by the value of the surface gravity of the metric. While the
metric can receive $\alpha'$ corrections, the relation between Hawking
temperature and surface gravity does not change. This is not the case for the
Bekenstein-Hawking entropy, which, in presence of $\alpha'$ corrections
(higher-order in curvature corrections in general) is no longer determined by
the area of the horizon which also receives $\alpha'$ corrections coming from
those of the metric. Based on previous work \cite{Lee:1990nz,Wald:1993nt}, in
Ref.~\cite{Iyer:1994ys} Iyer and Wald gave a prescription to derive an entropy
formula in diffeomorphism-invariant theories. The main fact that characterizes
this prescription is that the entropy computed using it satisfies the first
law of black-hole mechanics \cite{Bardeen:1973gs}.

Iyer and Wald's prescription is based on a series of assumptions about the
field content, which has to consist of tensor fields only. The only tensor
field in our current understanding of Nature is the metric,  the rest
being connections and sections of different gauge bundles or, in other words,
field with some kind of gauge freedom. The validity of Iyer and Wald's
prescription has subsequently extended to theories that include fields with
gauge freedoms in Refs.~\cite{Jacobson:2015uqa,Prabhu:2015vua,Aneesh:2020fcr},
but the Heterotic Superstring effective action (and many other string
effective actions) include a field which is not a connection or a section of
some gauge bundle: the Kalb-Ramond field. This complication has been ignored
in most of the string literature\footnote{An independent derivation of an
  entropy formula using Wald's formalism and dealing with some of the problems
  that the presence of the Kalb-Ramond field raises has been made in
  Ref.~\cite{Edelstein:2019wzg}. The final entropy formula derived there
  depends on a compensating gauge parameter which was left undetermined. This
  makes a comparison with the entropy formula we will derive impossible. For
  instance, it is not possible to compute the entropy of the Strominger-Vafa
  black hole using this formula, unless one can prove that the unknown term
  does not contribute to it. Although in that reference it is argued that, at
  least in certain relevant cases, this is indeed the case. In the same
  reference it is also shown that the invariance of their entropy formula
  under local Lorentz transformations depends on it, which seems
  contradictory.}  and the Iyer-Wald prescription has been naively applied
with results that seem to be compatible with the microscopic calculations of
the entropy.\footnote{Wald's formalism's first step consists in the proof of a
  first law of black-hole mechanics for the theory under consideration. A
  first law for the Heterotic Superstring effective action to first order in
  $\alpha'$ has not yet been proven, although it is widely assumed to exist
  (for instance, in the derivation of the entropy formula of
  Ref.~\cite{Edelstein:2019wzg}).}

For instance, in Ref.~\cite{Cano:2018qev}, the entropy of the (heterotic
version of the) $\alpha'$-corrected Strominger-Vafa black hole was computed
using the Iyer-Wald prescription directly in the 10-dimensional action. The
result obtained was compatible with that of the microscopic calculation
carried out in Ref.~\cite{Kraus:2005zm} to first-order in $\alpha'$, with an
appropriate identification between the charges carried by the black hole and
associated string background \cite{Faedo:2019xii}. More precisely, the entropy
obtained was interpreted in Ref.~\cite{Cano:2018qev} as the
$\mathcal{O}(\alpha')$ truncation of the expansion in powers of $\alpha'$ of
the exact result found in Ref.~\cite{Kraus:2005zm}.

This interpretation, however, was a bit puzzling, because in
Ref.~\cite{Cano:2018qev}, it was argued that the near-horizon region of the
black-hole solution, which determines the entropy, should not receive further
$\alpha'$ corrections.\footnote{The complete black-hole solution may receive
  further corrections.} Furthermore, an explicit calculation shows that at
least the $\mathcal{O}(\alpha^{\prime\, 2})$ corrections to the entropy vanish
identically \cite{kn:TOM}. All this suggests that the result obtained for the
entropy in Ref.~\cite{Cano:2018qev} should be exact to all orders in $\alpha'$
and, therefore, it should be identical to the result of the microscopic
calculation of Ref.~\cite{Kraus:2005zm}.

This puzzle was solved in Ref.~\cite{Faedo:2019xii}, where it was observed
that the dependence of the action on the Riemann curvature\footnote{According
  to the Iyer-Wald prescription, the entropy formula only depends on the
  occurrences of the Riemann tensor in the action.} in the Lorentz
Chern-Simons term of the Kalb-Ramond field strength is changed by dimensional
reduction. Taking into account this change, which amounts to a factor of 2
with respect to the result of Ref.~\cite{Cano:2018qev}, the macroscopic
entropy computed at first order in $\alpha'$ using naively the Iyer-Wald
formula matches the exact microscopic result. This gives further support to
the conjecture that the black-hole solution does not receive further $\alpha'$
corrections and may be considered an exact Heterotic Superstring solution.

The results of Ref.~\cite{Faedo:2019xii} made clear that, in the case of the
Heterotic Superstring effective action, the entropy formula has to be derived
from the dimensionally-reduced action in order to determine correctly the
dependence of the action of the lower-dimensional Riemann tensor. One of our
goals in this paper is to perform the dimensional reduction of the Heterotic
Superstring effective action to first order in $\alpha'$ over a circle to
apply to it the Iyer-Wald prescription and obtain an entropy formula. This
entropy formula can only be applied to $d$-dimensional black holes that can be
obtained by trivial compactification on T$^{9-d}$ and a non-trivial
compactification on a circle. For instance, it can be applied to the heterotic
version of the Strominger-Vafa black hole because it can be obtained from a
10-dimensional solution by trivial compactification on T$^{4}$, to 6
dimensions and a non-trivial compactification on a circle from 6 to 5
dimensions. It can also be applied to the non-supersymmetric 4-dimensional
Reissner-Nordstr\"om black hole of Ref.~\cite{Khuri:1995xq}, which can be
obtained from pure 5-dimensional gravity and, therefore, can be obtained from
a purely gravitational 10-dimensional solution by trivial compactification on
T$^{5}$ to 5 dimensions and, then, by a non-trivial compactification on a
circle from 5 to 4 dimensions. Actually, the entropy formula
Eq.~(\ref{eq:entropyformulastringframe}) that we are going to derive in
Section~\ref{sec-entropyformula} has been applied to a non-extremal version of
the 4-dimensional Reissner-Nordstr\"om black hole we just discussed, in
Ref.~\cite{Cano:2019ycn}. While the microscopic interpretation of the entropy
of this black hole is unknown, being a black hole with finite temperature, one
can check that the first law of thermodynamics is indeed satisfied because the
temperature computed from the $\alpha'$-corrected metric and the entropy
computed from the $\alpha'$-corrected metric with the $\alpha'$-corrected
entropy formula are related by the thermodynamic relation

\begin{equation}
\frac{\partial S}{\partial M}=\frac{1}{T}\, .
\end{equation}

This paper's second goal has to do with one of the most interesting and
characteristic properties of String Theory: T~duality.\footnote{For a review
  with many early references see Ref.~\cite{Alvarez:1994dn}.} T~duality
relates two string theories compactified in circles of dual radii. The spectra
of the two theories can be put into one-to-one correspondence and, from the
lower dimensional point of view, they are essentially identical, up to charge
identifications.\footnote{Charges related to Kaluza-Klein momentum and charges
  related to the winding number along the compact direction should be
  interchanged.} More generally, Buscher \cite{Buscher:1987sk,Buscher:1987qj}
showed that two string backgrounds with one isometry whose background fields
are related by the so-called \textit{Buscher T~duality rules} are equivalent.

String backgrounds related by T~duality may have very different geometries and
properties in spite of their stringy equivalence. On the other hand, T~duality
(via the Buscher rules) can be used actively to generate new (dual) string
backgrounds from known backgrounds. This is what makes this duality so
interesting.

Perhaps not surprisingly, the Buscher rules can be derived from the string
effective action: the dual\footnote{That is, with fields related by the
  Buscher rules.} Kaluza-Klein compactifications of two effective actions on a
circle give the same $(d-1)$-dimensional action and the same equations of
motion.  In practice, one can perform identical Kaluza-Klein
compactifications, determine the relation between the $(d-1)$-dimensional
fields of the two actions (which is usually very simple because it does not
involve the $(d-1)$-dimensional string metric or Kalb-Ramond field) and
rewrite this relation in terms of the components of the original
$d$-dimensional fields \cite{Bergshoeff:1994dg}. This relation is just the
Buscher T~duality rules. This strategy has been successfully used to find the
extension of the Buscher T~duality rules that relates equivalent type~IIA and
type~IIB superstring backgrounds \cite{Bergshoeff:1995as} and higher-rank
Ramond-Ramond potentials \cite{Meessen:1998qm}.

In the context of the Heterotic Superstring, this strategy was used in
\cite{Bergshoeff:1995cg} to find the first-order in $\alpha'$ corrections to
the Buscher rules.\footnote{At zeroth-order in $\alpha'$, the Heterotic
  Superstring effective action only describes the so-called \textit{common
    sector} of Neveu-Schwarz-Neveu-Schwarz fields, so the Buscher rules are
  just those found by Buscher.} Only the Yang-Mills fields were included at
order $\alpha'$, but, taking into account that the torsionful spin connection
enters the action in exactly the same way as the Yang-Mills fields
\cite{Bergshoeff:1988nn}, it was possible to find the $\alpha'$ corrections to
the Buscher rules.

The $\alpha'$-corrected Buscher rules are of no use if there are no
$\alpha'$-corrected solutions at one's disposal to generate new solutions or to check their equivalence. For this reason,
the results of Ref.~\cite{Bergshoeff:1995cg} were sleeping the ``sleep of the
just''\footnote{As a matter of fact, they have partially re-derived several
  times \cite{Serone:2005ge,Bedoya:2014pma}. Other studies of the effect of
  $\alpha'$ corrections on T~duality and O$(d,d)$ transformations in toroidal
  compactifications, sometimes in extended set-ups (such as Double Field
  Theory) can be found
  \cite{Meissner:1996sa,Kaloper:1997ux,Hohm:2014eba,Marques:2015vua,Baron:2017dvb,Eloy:2020dko}.}
until quite recently, when they were first applied to $\alpha'$-corrected
self-T-dual solutions, providing a highly non-trivial test of both the
$\alpha'$ corrections of the solutions and of the T~duality rules.

Our second goal will be to study the T~duality invariance of the complete
dimensionally-reduced Heterotic Superstring effective action and of the
entropy formula that follows from it. While the $\alpha'$-corrected Buscher
rules will be those of Ref.~\cite{Bergshoeff:1995cg}, the complete reduced
action will have many more $\mathcal{O}(\alpha')$ terms  than the action
obtained there. The invariance of the action under T~duality suggests that
they will contribute to the entropy in a T~duality-invariant form, and we will
prove that this is the case.\footnote{\label{foot:esa} It follows trivially
  from the invariance of the lower-dimensional string metric and dilaton under
  T~duality that the zeroth-order in $\alpha'$ temperature and entropy (the
  area) are also T~duality invariant. This property was proven by Horowitz and
  Welch in Ref.~\cite{Horowitz:1993wt} before the relation between the Buscher
  rules and dimensional reduction was established in
  Ref.~\cite{Bergshoeff:1994dg}. Recently, it has been investigated again from
  the same point of view in Refs.~\cite{Edelstein:2018ewc,Edelstein:2019wzg}
  to first order in $\alpha'$, but, again, the relation between dimensional
  reduction and T~duality and the invariance of the lower-dimensional string
  metric and dilaton field lead, trivially, to the invariance of the
  $\alpha'$-corrected temperature. The invariance of the action under
  T~duality at this order implies that of the entropy formula using the
  Iyer-Wald prescription because the Riemann curvature is T~duality
  invariant.}

This paper is organized as follows: we introduce the Heterotic Superstring
effective action to first order in $\alpha'$ following
Ref.~\cite{Bergshoeff:1989de} in Section~\ref{sec-heteroticalpha}. In
Section~\ref{sec-dimredO1}, we revisit the dimensional reduction on a circle of
the action at zeroth order in $\alpha'$ as a warm-up exercise and also because
we will need some of the results when we consider the higher-order terms in
Section~\ref{sec-dimredOalpha}. In that section we will obtain the complete
dimensionally-reduced action to first order in $\alpha'$, we will find the
T~duality rules and we will prove the invariance of the action under those
T~duality rules.  In Section~\ref{sec-entropyformula}, we will use the
dimensionally-reduced T~duality-invariant action to derive an entropy formula
using the Iyer-Wald prescription and we will apply it to the heterotic version
of the $\alpha'$-corrected Strominger-Vafa black hole of
Ref.~\cite{Cano:2018qev}.  We will end by discussing our results and future
work on these topics in Section~\ref{sec-discussion}.

%%%%%%%%%%%%%%%%%%%%%%%%%%%%%%%%%%%%%%%%%%%%%%%%%%%%%%%%%%%%%%%%%%%%%%
%%%%%%%%%%%%%%%%%%%%%%%%%%%%%%%%%%%%%%%%%%%%%%%%%%%%%%%%%%%%%%%%%%%%%%
%%%%%%%%%%%%%%%%%%%%%%%%%%%%%%%%%%%%%%%%%%%%%%%%%%%%%%%%%%%%%%%%%%%%%%
%%%%%%%%%%%%%%%%%%%%%%%%%%%%%%%%%%%%%%%%%%%%%%%%%%%%%%%%%%%%%%%%%%%%%%
\section{The Heterotic Superstring effective action to
  \texorpdfstring{$\mathcal{O}(\alpha')$}{O(α')}}
\label{sec-heteroticalpha}
%%%%%%%%%%%%%%%%%%%%%%%%%%%%%%%%%%%%%%%%%%%%%%%%%%%%%%%%%%%%%%%%%%%%%%
%%%%%%%%%%%%%%%%%%%%%%%%%%%%%%%%%%%%%%%%%%%%%%%%%%%%%%%%%%%%%%%%%%%%%%
%%%%%%%%%%%%%%%%%%%%%%%%%%%%%%%%%%%%%%%%%%%%%%%%%%%%%%%%%%%%%%%%%%%%%%
%%%%%%%%%%%%%%%%%%%%%%%%%%%%%%%%%%%%%%%%%%%%%%%%%%%%%%%%%%%%%%%%%%%%%%

Let us start by reviewing the Heterotic Superstring effective action to
$\mathcal{O}(\alpha')$. We will use the formulation given in
Ref.~\cite{Bergshoeff:1989de}, but written in the conventions of
Ref.~\cite{Ortin:2015hya}.\footnote{The relation with the fields in
  Ref.~\cite{Bergshoeff:1989de} can be found in
  Ref.~\cite{Fontanella:2019avn}.} In this formulation, the action is
constructed recursively order by order in $\alpha'$.

The zeroth-order 3-form field strength of the Kalb-Ramond 2-form $B$
is defined as

\begin{equation}
H^{(0)}{}_{\mu\nu\rho} \equiv 3\partial_{[\mu}B_{\nu\rho]}\, ,
\end{equation}

\noindent
and it contributes as torsion to the zeroth-order torsionful spin
connections

\begin{equation}
{\Omega}^{(0)}_{(\pm)\, \mu}{}^{{a}}{}_{{b}} 
=
{\omega}_{\mu}{}^{{a}}{}_{{b}}
\pm
\tfrac{1}{2}{H}^{(0)}{}_{{\mu}}{}^{{a}}{}_{{b}}\, ,
\end{equation}

\noindent
where ${\omega}_{\mu}{}^{{a}}{}_{{b}}$ is the (torsionless, metric-compatible)
Levi-Civita spin connection 1-form.

The corresponding zeroth-order Lorentz curvature 2-forms and
Chern-Simons 3-forms are defined as

\begin{eqnarray}
  \label{eq:R0def}
{R}^{(0)}_{(\pm)\, \mu\nu}{}^{{a}}{}_{{b}}
& = & 
2\partial_{[\mu|} {\Omega}^{(0)}_{(\pm)\, |\nu]}{}^{{a}}{}_{{b}}
-2 {\Omega}^{(0)}_{(\pm)\, [\mu|}{}^{{a}}{}_{{c}}\,
{\Omega}^{(0)}_{(\pm)\, |\nu]}{}^{{c}}{}_{{b}}\, ,
\\
  & & \nonumber \\
  \label{eq:oL0def}
{\omega}^{{\rm L}\, (0)}_{(\pm)}
& = &  
3{R}^{ (0)}_{(\pm)\, [\mu\nu|}{}^{{a}}{}_{{b}} 
{\Omega}^{ (0)}_{(\pm)\, |\rho]}{}^{{b}}{}_{{a}} 
+2
{\Omega}^{ (0)}_{(\pm)\, [\mu|}{}^{{a}}{}_{{b}} \,
{\Omega}^{ (0)}_{(\pm)\, |\nu|}{}^{{b}}{}_{{c}} \,
{\Omega}^{ (0)}_{(\pm)\, |\rho]}{}^{{c}}{}_{{a}}\, .  
\end{eqnarray}

The gauge field 1-form is $A^{A}{}_{\mu}$, where $A,B,C,\ldots$ are the
adjoint gauge indices of some group that we will not specify. The gauge field
strength and the Chern-Simons 3-forms are defined by

\begin{eqnarray}
{F}^{A}{}_{\mu\nu}
& = & 
2\partial_{[\mu}{A}^{A}{}_{\nu]}+f_{BC}{}^{A}{A}^{B}{}_{[\mu}{A}^{C}{}_{\nu]}\, , 
\\
& & \nonumber \\
{\omega}^{\rm YM}
& = & 
3F_{A\, [\mu\nu}{A}^{A}{}_{\rho]}
-f_{ABC}{A}^{A}{}_{[\mu}{A}^{B}{}_{\nu}{A}^{C}{}_{\rho]}\, ,
\end{eqnarray}

\noindent
where we have lowered the adjoint group indices using the Killing metric of
$K_{AB}$: $f_{ABC}\equiv f_{AB}{}^{D}K_{DC}$ and of the gauge fields
$F_{A\, \mu\nu}\equiv K_{AB}F^{B}{}_{\mu\nu}$.

Then, at first order

\begin{eqnarray}
H^{(1)}{}_{\mu\nu\rho}
& = & 
3\partial_{[\mu}B_{\nu\rho]}
+\frac{\alpha'}{4}\left({\omega}^{\rm YM}{}_{\mu\nu\rho}
+{\omega}^{{\rm L}\, (0)}_{(-)\, \mu\nu\rho}\right)\, ,  
\\
& & \nonumber \\
{\Omega}^{(1)}_{(\pm)\, \mu}{}^{{a}}{}_{{b}} 
& = & 
{\omega}_{\mu}{}^{{a}}{}_{{b}}
\pm
\tfrac{1}{2}{H}^{(1)}_{{\mu}}{}^{{a}}{}_{{b}}\, ,
\\
& & \nonumber \\
{R}^{(1)}_{(\pm)\, \mu\nu}{}^{{a}}{}_{{b}}
& = & 
2\partial_{[\mu|} {\Omega}^{(1)}_{(\pm)\, |\nu]}{}^{{a}}{}_{{b}}
-2 {\Omega}^{(1)}_{(\pm)\, [\mu|}{}^{{a}}{}_{{c}}\,
{\Omega}^{(1)}_{(\pm)\, |\nu]}{}^{{c}}{}_{{b}}\, ,
\\
& & \nonumber \\
{\omega}^{{\rm L}\, (1)}_{(\pm)\, \mu\nu\rho}
& = &  
3{R}^{ (1)}_{(\pm)\, [\mu\nu|}{}^{{a}}{}_{{b}} 
{\Omega}^{ (1)}_{(\pm)\, |\rho]}{}^{{b}}{}_{{a}} 
+2
{\Omega}^{(1)}_{(\pm)\, [\mu|}{}^{{a}}{}_{{b}}\, 
{\Omega}^{(1)}_{(\pm)\, |\nu|}{}^{{b}}{}_{{c}}\, 
{\Omega}^{(1)}_{(\pm)\, |\rho]}{}^{{c}}{}_{{a}}\, .  
 \\
& & \nonumber \\
H^{(2)}{}_{\mu\nu\rho}
& = & 
3\partial_{[\mu}B_{\nu\rho]}
+\frac{\alpha'}{4}\left({\omega}^{\rm YM}{}_{\mu\nu\rho}
+{\omega}^{{\rm L}\, (1)}_{(-)\, \mu\nu\rho}\right)\, ,  
\end{eqnarray}

\noindent
etc.

Only
$\Omega^{(0)}_{(\pm)\, \mu},{R}^{(0)}_{(\pm)\, \mu\nu}{}^{a}{}_{b},
\omega^{{\rm L}\, (0)}_{(\pm)\, \mu\nu\rho}$ and $ H^{(1)}{}_{\mu\nu\rho}$
(plus the Yang-Mills fields) occur in the action.  In practice, though, it is
more convenient to work with the higher-order objects, neglecting the terms of
higher order in $\alpha'$ when necessary. Thus, from now on we will suppress
the $(n)$ upper indices when they do not play a relevant role.

In terms of all these objects, the Heterotic Superstring effective action in
the string frame and to first-order in $\alpha'$ can be written as

\begin{equation}
\label{heterotic}
{S}
=
\frac{g_{s}^{2}}{16\pi G_{N}^{(10)}}
\int d^{10}x\sqrt{|{g}|}\, 
e^{-2{\phi}}\, 
\left\{
{R} 
-4(\partial{\phi})^{2}
+\tfrac{1}{12}{H}^{2}
-\dfrac{\alpha'}{8}\left[
{F}_{A}\cdot {F}^{A}
+
{R}_{(-)}{}^{{a}}{}_{{b}}\cdot
{R}_{(-)}{}^{{b}}{}_{{a}}
\right]
\right\}\, ,
\end{equation}

\noindent
where $G_{N}^{(10)}$ is the 10-dimensional Newton constant, $\phi$ is the
dilaton field, the vacuum expectation value of $e^{\phi}$ is the Heterotic
Superstring coupling constant $g_{s}$, $R$ is the Ricci scalar of the
string-frame metric $g_{\mu\nu}$ and the dot indicates the contraction of the
indices of 2-forms:
${F}_{A}\cdot {F}^{A}\equiv {F}_{A\, \mu\nu}{F}^{A\, \mu\nu}$.

%%%%%%%%%%%%%%%%%%%%%%%%%%%%%%%%%%%%%%%%%%%%%%%%%%%%%%%%%%%%%%%%%%%%%%
%%%%%%%%%%%%%%%%%%%%%%%%%%%%%%%%%%%%%%%%%%%%%%%%%%%%%%%%%%%%%%%%%%%%%%
%%%%%%%%%%%%%%%%%%%%%%%%%%%%%%%%%%%%%%%%%%%%%%%%%%%%%%%%%%%%%%%%%%%%%%
%%%%%%%%%%%%%%%%%%%%%%%%%%%%%%%%%%%%%%%%%%%%%%%%%%%%%%%%%%%%%%%%%%%%%%
\section{Dimensional reduction on S$^{1}$ at zeroth order in $\alpha'$}
\label{sec-dimredO1}
%%%%%%%%%%%%%%%%%%%%%%%%%%%%%%%%%%%%%%%%%%%%%%%%%%%%%%%%%%%%%%%%%%%%%% 
%%%%%%%%%%%%%%%%%%%%%%%%%%%%%%%%%%%%%%%%%%%%%%%%%%%%%%%%%%%%%%%%%%%%%%
%%%%%%%%%%%%%%%%%%%%%%%%%%%%%%%%%%%%%%%%%%%%%%%%%%%%%%%%%%%%%%%%%%%%%%
%%%%%%%%%%%%%%%%%%%%%%%%%%%%%%%%%%%%%%%%%%%%%%%%%%%%%%%%%%%%%%%%%%%%%%

As a warm-up exercise (and also because of the recursive definition of the
action that will make necessary the zeroth-order fields in the first-order
action), we review the well-known dimensional reduction of the action at
zeroth order in $\alpha'$ using the Scherk-Schwarz formalism
\cite{Scherk:1979zr}. We add hats to all the 10-dimensional objects (fields,
indices, coordinates) and split the 10-dimensional world indices as
$(\hat{\mu})=(\mu,\underline{z})$ and the 10-dimensional indices as
$(\hat{a})=(a,z)$.

The Zehnbein and inverse-Zehnbein components $\hat{e}_{\hat{\mu}}{}^{\hat{a}}$
and $\hat{e}_{\hat{a}}{}^{\hat{\mu}}$ can be put in an upper-triangular form
by a local Lorentz transformation and, then, they can be decomposed in terms
of the 9-dimensional Vielbein and inverse Vielbein components
$e_{\mu}{}^{a},e_{a}{}^{\mu}$, Kaluza-Klein (KK) vector $A_{\mu}$ and KK
scalar $k$ as

\begin{equation}
\label{eq:standardVielbeinAnsatz}
\left( \hat{e}_{\hat{\mu}}{}^{\hat{a}} \right) = 
\left(
\begin{array}{c@{\quad}c}
e_{\mu}{}^{a} & kA_{\mu} \\
&\\[-3pt]
0       & k    \\
\end{array}
\right)\!, 
\hspace{1cm}
\left(\hat{e}_{\hat{a}}{}^{\hat{\mu}} \right) =
\left(
\begin{array}{c@{\quad}c}
e_{a}{}^{\mu} & -A_{a}  \\
& \\[-3pt]
0       & k^{-1} \\
\end{array}
\right)\!,
\end{equation}

\noindent
where $A_{a}= e_{a}{}^{\mu} A_{\mu}$. We will always assume that all the
9-dimensional fields with Lorentz indices are 9-dimensional world tensors
contracted with the 9-dimensional Vielbeins. For instance, the KK fields
strength $F_{ab}$ is

\begin{equation}
\label{eq:KKfieldstrength}
F_{ab} = e_{a}{}^{\mu} e_{b}{}^{\nu} F_{\mu\nu},
\hspace{1cm}
F_{\mu\nu} \equiv 2\partial_{[\mu}A_{\nu]},
\end{equation}

The components of the 10-dimensional spin connection
$\hat{\omega}_{\hat{a}\hat{b}\hat{c}}$ decompose into those of the
9-dimensional one $\omega_{abc}$ and $F_{ab}$ as

\begin{equation}
\label{eq:standardspinconnectionreduction}
\begin{array}{rclrcl}
\hat{\omega}_{abc} & = & \omega_{abc}, &
\hat{\omega}_{abz} & = & \frac{1}{2} k F_{ab},
\\
& & & & & \\
\hat{\omega}_{zbc} & = & -\frac{1}{2} k F_{bc},
\hspace{1.5cm}& 
\hat{\omega}_{zbz} & = & -\partial_{b} \ln{k}.\\
\end{array}
\end{equation}

Then, using the Palatini identity, it is not difficult to see that the first
two terms in the action Eq.~(\ref{heterotic}) take the following 9-dimensional
form (up to a total derivative):

\begin{equation}
  \begin{aligned}
    \int d^{10}\hat{x}\sqrt{|\hat{g}|}\, e^{-2\hat{\phi}}\, \left\{ \hat{R}
      -4(\partial\hat{\phi})^{2} \right\}
        & = \\
        & \\
        & \hspace{-3cm}
        \int dz\int d^{9}x\sqrt{|g|}\,
        e^{-2\phi}\, \left\{ R -4(\partial\phi)^{2} +(\partial \log{k})^{2}
          -\tfrac{1}{4}k^{2}F^{2} \right\}\, ,
          \end{aligned}
\end{equation}

\noindent
where the 9-dimensional dilaton field is related to the 10-dimensional one by

\begin{equation}
\phi \equiv \hat{\phi} -\tfrac{1}{2}\log{k}\, .  
\end{equation}

At zeroth order in $\alpha'$, the last term that we have to reduce is the
kinetic term of the Kalb-Ramond 2-form $\sim \hat{H}^{(0)\,2}$. Following
Scherk and Schwarz, we consider the Lorentz components of the 3-form field
strength, because they are automatically gauge-invariant combinations. The
$\hat{H}^{(0)}{}_{abz}$ components give

\begin{equation}
  \hat{H}^{(0)}{}_{abz}
  =
  k^{-1}e_{a}{}^{\mu}e_{b}{}^{\nu} \hat{H}^{(0)}{}_{\mu\nu\underline{z}} 
 =
 k^{-1}e_{a}{}^{\mu}e_{b}{}^{\nu} 2\partial_{[\mu} \hat{B}_{\nu]\underline{z}}\, .
\end{equation}

It is, then, appropriate to define the zeroth-order ``winding''\footnote{This
  vector couples electrically to the string modes with non-vanishing winding
  numbers, just as the KK vector field couples to those with non-vanishing
  momentum in the internal direction.} vector field $B^{(0)}{}_{\mu}$ and its
field strength $G^{(0)}{}_{\mu\nu}$ by

\begin{equation}
  B^{(0)}{}_{\mu} \equiv \hat{B}_{\mu\underline{z}}\, ,
  \hspace{1cm}
  G^{(0)}{}_{\mu\nu} \equiv 2\partial_{[\mu}B^{(0)}{}_{\nu]}\, ,
\end{equation}

\noindent
so that 

\begin{equation}
  \hat{H}^{(0)}{}_{abz}
  =
  k^{-1}G^{(0)}{}_{ab}\, .
\end{equation}

The second gauge-invariant combination is

\begin{equation}
  \hat{H}^{(0)}{}_{abc}
  =
  e_{a}{}^{\mu}e_{b}{}^{\nu}e_{c}{}^{\rho}
  \left(
    \hat{H}^{(0)}{}_{\mu\nu\rho} -3 A_{[\mu}\hat{H}^{(0)}{}_{\nu\rho]\underline{z}}
  \right)\, ,
\end{equation}

\noindent
which suggests the definition

\begin{equation}
  H^{(0)}{}_{\mu\nu\rho}
  \equiv
  \hat{H}^{(0)}{}_{\mu\nu\rho} -3A_{[\mu}\hat{H}^{(0)}{}_{\nu\rho]\underline{z}}
= 3\partial_{[\mu}\hat{B}_{\nu\rho]} -6A_{[\mu}\partial_{\nu}
B^{(0)}{}_{\rho]}\, .
\end{equation}

We could simply identify $\hat{B}_{\nu\rho}$ with the 9-dimensional
Kalb-Ramond field, but it is customary (and convenient) to use a
T~duality-invariant definition. T~duality will interchange KK momentum and
winding, and therefore, will interchange $A_{\mu}$ with $B^{(0)}{}_{\mu}$,
modifying the Chern-Simons term in the above form of $H_{\mu\nu\rho}$. We can,
however, rewrite it in the form

\begin{equation}
  H^{(0)}{}_{\mu\nu\rho}
  = 3\partial_{[\mu}\left(\hat{B}_{\nu\rho]} +A_{|\nu}B^{(0)}{}_{\rho]}\right)
  -\tfrac{3}{2}A_{[\mu}G^{(0)}{}_{\nu\rho]}-\tfrac{3}{2}B^{(0)}{}_{[\mu}F_{\nu\rho]} \, ,
\end{equation}

\noindent
and identify the T~duality-invariant 9-dimensional Kalb-Ramond 2-form

\begin{equation}
B^{(0)}{}_{\mu\nu} \equiv \hat{B}_{\mu\nu} +A_{[\mu}B^{(0)}{}_{\nu]}\, ,  
\end{equation}

\noindent
with the final result

\begin{equation}
  H^{(0)}{}_{\mu\nu\rho}
  = 3\partial_{[\mu}\hat{B}^{(0)}{}_{\nu\rho]}
  -\tfrac{3}{2}A_{[\mu}G^{(0)}{}_{\nu\rho]}-\tfrac{3}{2}B^{(0)}{}_{[\mu}F_{\nu\rho]}\, .
\end{equation}

Then, after integrating over the length of the compact coordinate $z$
($2\pi\ell_{s}$ by convention) the 9-dimensional action to zeroth order in
$\alpha'$ takes the form

\begin{equation}
  \label{eq:heterotic9order0}
  S
  =
  \frac{g_{s}^{2}(2\pi\ell_{s})}{16\pi G_{N}^{(10)}}
  \int d^{9}x\sqrt{|g|}\,
  e^{-2\phi}\, \left\{ R -4(\partial\phi)^{2} +(\partial \log{k})^{2}
    -\tfrac{1}{4}k^{2}F^{2}  -\tfrac{1}{4}k^{-2}G^{(0)\, 2}
    +\tfrac{1}{12}H^{(0)\, 2}\right\}\, .
\end{equation}

This action is invariant under the T~duality transformations

\begin{equation}
  \label{eq:0thorderTduality}
  A_{\mu}'
  = B^{(0)}{}_{\mu}\, , 
  \hspace{1cm}
  B^{(0)}{}_{\mu}'
  = 
  A_{\mu}\, ,
  \hspace{1cm}
  k'
   = 
  1/k\, .
\end{equation}

Taking into account the relations between the 10- and 9-dimensional fields,
collected in Appendix~\ref{sec-10versus9atorder0}, it is easy to see that the
above T~duality transformations correspond to the following transformations of
the 10-dimensional fields known as \textit{Buscher rules}
\cite{Buscher:1987sk,Buscher:1987qj}:

\begin{equation}
\label{eq:Buscherrules}
\begin{array}{rclrcl}
\hat{g}^{\prime}_{\underline{z}\underline{z}} & = &
1/\hat{g}_{\underline{z}\underline{z}}\, , &
\hat{B}^{\prime}_{\mu \underline{z}} & = &
\hat{g}_{\mu \underline{z}}/\hat{g}_{\underline{z}\underline{z}}\, ,
\\
& & & & &
\\
\hat{g}^{\prime}_{\mu \underline{z}} & = &
\hat{B}_{\mu \underline{z}}/\hat{g}_{\underline{z}\underline{z}}\, , &
\hat{B}^{\prime}_{\mu\nu} & = &
\hat{B}_{\mu\nu}+2\hat{g}_{[\mu|\underline{z}|}
\hat{B}_{\nu] \underline{z}}/\hat{g}_{\underline{z}\underline{z}}\, ,
\\
& & & & &
\\
\hat{g}^{\prime}_{\mu\nu} & = &
\hat{g}_{\mu\nu}-(\hat{g}_{\mu \underline{z}}\hat{g}_{\nu \underline{z}}-
\hat{B}_{\mu \underline{z}}\hat{B}_{\nu \underline{z}})
/\hat{g}_{\underline{z}\underline{z}}\, , &
\hat{\phi}^{\prime} & = & \hat{\phi} -\frac{1}{2}\ln
|\hat{g}_{\underline{z}\underline{z}}|\, .
\end{array}
\end{equation}

%%%%%%%%%%%%%%%%%%%%%%%%%%%%%%%%%%%%%%%%%%%%%%%%%%%%%%%%%%%%%%%%%%%%%%
%%%%%%%%%%%%%%%%%%%%%%%%%%%%%%%%%%%%%%%%%%%%%%%%%%%%%%%%%%%%%%%%%%%%%%
%%%%%%%%%%%%%%%%%%%%%%%%%%%%%%%%%%%%%%%%%%%%%%%%%%%%%%%%%%%%%%%%%%%%%%
%%%%%%%%%%%%%%%%%%%%%%%%%%%%%%%%%%%%%%%%%%%%%%%%%%%%%%%%%%%%%%%%%%%%%%
\section{Dimensional reduction on S$^{1}$ at $\mathcal{O}(\alpha'$)}
\label{sec-dimredOalpha}
%%%%%%%%%%%%%%%%%%%%%%%%%%%%%%%%%%%%%%%%%%%%%%%%%%%%%%%%%%%%%%%%%%%%%% 
%%%%%%%%%%%%%%%%%%%%%%%%%%%%%%%%%%%%%%%%%%%%%%%%%%%%%%%%%%%%%%%%%%%%%%
%%%%%%%%%%%%%%%%%%%%%%%%%%%%%%%%%%%%%%%%%%%%%%%%%%%%%%%%%%%%%%%%%%%%%%
%%%%%%%%%%%%%%%%%%%%%%%%%%%%%%%%%%%%%%%%%%%%%%%%%%%%%%%%%%%%%%%%%%%%%%

The reduction of the first two terms in the effective action is not modified
by the inclusion of $\alpha'$ corrections. The definitions of 9-dimensional
metric, dilaton and KK vector and scalar in terms of the 10-dimensional fields
are not modified by them either.  We expect modifications in the definitions
of the 9-dimensional Kalb-Ramond 2-form and of the winding vector, though,
because of the presence of the Lorentz and Yang-Mills Chern-Simons 3-forms in
$\hat{H}^{(1)}$.

It is convenient to start by studying the dimensional reduction of the
Yang-Mills fields. The Lorentz-indices decomposition of the gauge field is

\begin{subequations}
  \begin{align}
    \hat{A}^{A}{}_{z}
    & =
      k^{-1}\hat{A}^{A}{}_{\underline{z}}\, ,
      \\
    & \nonumber \\
    \hat{A}^{A}{}_{a}
    & =
      e_{a}{}^{\mu}( \hat{A}^{A}{}_{\mu}- \hat{A}^{A}{}_{\underline{z}}A_{\mu})\, ,
  \end{align}
\end{subequations}

\noindent
which leads to the definition of the 9-dimensional adjoint scalars $\phi^{A}$
and gauge vectors

\begin{subequations}
  \begin{align}
    \varphi^{A}
    & \equiv
      k^{-1}\hat{A}^{A}{}_{\underline{z}}\, ,
    \\
    & \nonumber \\
    A^{A}{}_{\mu}
    & \equiv
      \hat{A}^{A}{}_{\mu}- \hat{A}^{A}{}_{\underline{z}}A_{\mu}\, .
  \end{align}
\end{subequations}

In terms of these variables, it is not difficult to see that the
components of 10-dimensional gauge field strength are given by

\begin{subequations}
  \label{eq:YMreduction}
  \begin{align}
    \hat{F}^{A}{}_{az}
    & =
      \mathfrak{D}_{a}\varphi^{A} +\varphi^{A}\partial_{a}\log{k}\, ,
    \\
    & \nonumber \\
    \hat{F}^{A}{}_{ab}
    & =
      F^{A}{}_{ab} +k\varphi^{A}F_{ab}\, ,
  \end{align}
\end{subequations}

\noindent
where $ F^{A}{}_{\mu\nu}$ is the standard Yang-Mills gauge field strength for
the 9-dimensional gauge fields $A^{A}{}_{\mu}$.

The reduction of the first, second and fourth terms in the action
Eq.~(\ref{heterotic}) gives (up to a total derivative)

\begin{equation}
  \begin{aligned}
    \int dz\int d^{9}x \sqrt{|g|}e^{-2\phi} \left\{ R -4(\partial\phi)^{2}
      +\left(1+\frac{\alpha'}{4}\varphi^{2}\right)(\partial
      \log{k})^{2}
      +\frac{\alpha'}{4}(\mathfrak{D}\varphi)^{2}
      +\frac{\alpha'}{4}\partial_{a}\log{k}\partial^{a}\varphi^{2}
      \right.
      & \\
      & \\
      \left.
      -\tfrac{1}{4}\left(1+\frac{\alpha'}{2}\varphi^{2}\right)k^{2}F^{2}
      -\frac{\alpha'}{8}\left(F_{A}\cdot F^{A}
      +2\varphi_{A}F^{A}\cdot  kF\right) \right\}\, ,
  &
  \end{aligned}
\end{equation}

\noindent
where $\varphi^{2}\equiv \varphi_{A}\varphi^{A}$,
$\mathfrak{D}_{\mu}\varphi^{A}= \partial_{\mu}\varphi^{A}
+f_{BC}{}^{A}A^{B}{}_{\mu}\varphi^{C}$ etc.

Let us now consider the reduction of the Kalb-Ramond 3-form field strength
$\hat{H}^{(1)}$, starting with the gauge-invariant combination

\begin{equation}
  \hat{H}^{(1)}{}_{abz}
  =
  k^{-1}e_{a}{}^{\mu}e_{b}{}^{\nu} \hat{H}^{(1)}{}_{\mu\nu\underline{z}} 
 =
 k^{-1}e_{a}{}^{\mu}e_{b}{}^{\nu}
 \left\{ 2\partial_{[\mu} \hat{B}_{\nu]\underline{z}}
   +\frac{\alpha'}{4}\left(\hat{\omega}^{\rm YM}{}_{\mu\nu\underline{z}}
     +\hat{\omega}^{\rm L (0)}_{(-)\, \mu\nu\underline{z}}\right)
   \right\}\, .
\end{equation}

\noindent
Using the above results for the Yang-Mills fields we find that

\begin{equation}
  \label{eq:oYMmnz}
  \hat{\omega}^{\rm YM}{}_{\mu\nu\underline{z}}
  =
  k\varphi_{A}\left(2F^{A}{}_{\mu\nu} +\varphi^{A}kF_{\mu\nu} \right)
  -2\partial_{[\mu|}\left(k\varphi_{A}A^{A}_{|\nu]} \right)\, .
\end{equation}

\noindent
The last term is a total derivative that can be absorbed into the definition
of the 9-dimensional vector field $B^{(1)}{}_{\mu}$ and the remaining terms
are manifestly gauge-invariant 2-forms.

We can use this result in the reduction of the Lorentz Chern-Simons 3-form;
after all, the only difference with the Yang-Mills Chern-Simons 3-form is the
gauge group, which now is the 10-dimensional Lorentz group. This is,
nevertheless, an important difference because this group is broken down to the
9-dimensional Lorentz group times U$(1)$ and we will have to take this fact
into account in a second step.

In order to profit from the previous result, we introduce the following
notation

\begin{subequations}
  \begin{align}
    \hat{A}^{\hat{a}\hat{b}}{}_{\hat{\mu}}
    & \equiv
      \hat{\Omega}^{(0)}_{(-)\, \hat{\mu}}{}^{\hat{a}\hat{b}}\, ,
    \\
    & \nonumber \\
    \hat{F}^{\hat{a}\hat{b}}{}_{\hat{\mu}\hat{\nu}}
    & \equiv
      \hat{R}^{(0)}_{(-)\, \hat{\mu}\hat{\nu}}{}^{\hat{a}\hat{b}}\, .
  \end{align}
\end{subequations}

\noindent
Then, a straightforward application of Eq.~(\ref{eq:oYMmnz}) gives

\begin{equation}
  \hat{\omega}^{\rm L (0)}_{(-)\, \mu\nu\underline{z}}
  =
  k\varphi^{\hat{a}}{}_{\hat{b}}\left(2F^{\hat{b}}{}_{\hat{a}\, \mu\nu}
    +\varphi^{\hat{b}}{}_{\hat{a}}kF_{\mu\nu} \right)
  -2\partial_{[\mu|}\left(k\varphi^{\hat{a}}{}_{\hat{b}}A^{\hat{b}}{}_{\hat{a}\,
      |\nu]} \right)\, ,
\end{equation}

\noindent
where

\begin{subequations}
  \label{eq:lorentzgaugefields}
  \begin{align}
    \varphi^{\hat{a}\hat{b}}
    & =
      k^{-1}\hat{\Omega}^{(0)}_{(-)\, \underline{z}}{}^{\hat{a}\hat{b}}\, ,
    \\
    & \nonumber \\
    A^{\hat{a}\hat{b}}{}_{\mu}
    & =
      \hat{\Omega}^{(0)}_{(-)\, \mu}{}^{\hat{a}\hat{b}}
      -A_{\mu}\hat{\Omega}^{(0)}_{(-)\, \underline{z}}{}^{\hat{a}\hat{b}}\, ,
  \end{align}
\end{subequations}

\noindent
and where $F^{\hat{a}\hat{b}}{}_{\mu\nu}$ is the standard field strength of
the gauge field $A^{\hat{a}\hat{b}}{}_{\mu}$ defined above. Decomposing now
the Lorentz indices, we obtain

\begin{equation}
  \label{eq:oL0decomposed}
  \begin{aligned}
    \hat{\omega}^{\rm L (0)}_{(-)\, \mu\nu\underline{z}}
    & =
    k\varphi^{a}{}_{b}\left(2F^{b}{}_{a\, \mu\nu}
      +\varphi^{b}{}_{a}kF_{\mu\nu} \right)
    +2k\varphi^{az}\left(2F_{a}{}^{z}{}_{\mu\nu}
      +\varphi_{a}{}^{z}kF_{\mu\nu} \right)
    \\
    & \\
    &
    -2\partial_{[\mu|}\left[k\left(\varphi^{\hat{a}}{}_{\hat{b}}
        A^{\hat{b}}{}_{\hat{a}\,|\nu]}\right) \right]\, . 
  \end{aligned}
\end{equation}

The components of these fields are

\begin{subequations}
  \begin{align}
    \varphi^{ab}
    & =
      -\tfrac{1}{2}\left(kF^{ab}+k^{-1}G^{(0)\, ab}\right)\, ,
    \\
    & \nonumber \\
    \varphi^{az}
    & =
      \partial^{a}\log{k}\, ,
    \\
    & \nonumber \\
    A^{ab}{}_{\mu}
    & =
      \omega_{\mu}{}^{ab}-\tfrac{1}{2}H^{(0)}{}_{\mu}{}^{ab}
      \equiv
      \Omega^{(0)}_{(-)\, \mu}{}^{ab}\, ,
    \\
    & \nonumber \\
    A^{az}{}_{\mu}
    & =
      -\tfrac{1}{2}\left(kF_{\mu}{}^{a}-k^{-1}G^{(0)}{}_{\mu}{}^{a}\right)\, ,
    \\
    & \nonumber \\
    F^{ab}{}_{\mu\nu}
    & =
    R^{(0)}_{(-)\, \mu\nu}{}^{ab}
      -\tfrac{1}{2}\left(kF_{[\mu}{}^{a}-k^{-1}G^{(0)}{}_{[\mu}{}^{a}\right)
      \left(kF_{\nu]}{}^{b}-k^{-1}G^{(0)}{}_{\nu]}{}^{b}\right)\, ,  
    \\
    & \nonumber \\
    F^{az}{}_{\mu\nu}
    & =
      -\mathcal{D}^{(0)}_{(-)\, [\mu}\left(kF_{\nu]}{}^{a}
      -k^{-1}G^{(0)}{}_{\nu]}{}^{a}\right)\, ,
  \end{align}
\end{subequations}

\noindent
where $ R^{(0)}_{(-)\, \mu\nu}{}^{ab}$ is the standard Lorentz curvature and
$\mathcal{D}^{(0)}_{(-)\, \mu}$ is the standard Lorentz-covariant derivative
with respect to the 9-dimensional torsionful spin connection
$\Omega^{(0)}_{(-)\, \mu}{}^{ab}$.

Replacing the above expressions in Eq.~(\ref{eq:oL0decomposed}) we obtain

\begin{equation}
  \begin{aligned}
    \hat{\omega}^{\rm L (0)}_{(-)\, \mu\nu\underline{z}}
    & =
-\tfrac{1}{2}k\left(kF^{a}{}_{b}+k^{-1}G^{(0)\, a}{}_{b}\right)
\left\{
 2 R^{(0)}_{(-)\, \mu\nu}{}^{b}{}_{a}
      -\left(kF_{[\mu|}{}^{b}-k^{-1}G^{(0)}{}_{[\mu}{}^{b}\right)
      \left(kF_{\nu]\, a}-k^{-1}G^{(0)}{}_{|\nu]\, a}\right)
    \right.
    \\
    & \\
    & \hspace{.5cm}
    \left.
-\tfrac{1}{2}\left(kF^{b}{}_{a}+k^{-1}G^{(0)\, b}{}_{a}\right)kF_{\mu\nu} \right\}
    -2\partial^{a}k\left[2\mathcal{D}^{(0)}_{(-)\, [\mu}\left(kF_{\nu]\, a}
      -k^{-1}G^{(0)}{}_{|\nu]\, a}\right)
      -\partial_{a}k F_{\mu\nu} \right]
    \\
    & \\
    & \hspace{.5cm}
    -2\partial_{[\mu|}\left[k\left(\varphi^{\hat{a}}{}_{\hat{b}}
        A^{\hat{b}}{}_{\hat{a}\,|\nu]}\right) \right]\, , 
  \end{aligned}
\end{equation}

\noindent
and

\begin{equation}
  \begin{aligned}
  \hat{H}^{(1)}{}_{cdz}
  & =
  k^{-1}e_{c}{}^{\mu}e_{d}{}^{\nu}
\left\{ 2\partial_{[\mu}\left[ \hat{B}_{\nu]\underline{z}}
    -\frac{\alpha'}{4}k\left(\varphi_{A}A^{A}{}_{|\nu]}+\varphi^{\hat{a}}{}_{\hat{b}}
        A^{\hat{b}}{}_{\hat{a}\,|\nu]}\right)
    \right]
    \right.
    \\
    & \\
    & \hspace{.5cm}
  +\frac{\alpha'}{4}
  k\varphi_{A}\left(2F^{A}{}_{\mu\nu} +\varphi^{A}kF_{\mu\nu} \right)
\\
& \\
& \hspace{.5cm}
-\tfrac{1}{2}k\left(kF^{a}{}_{b}+k^{-1}G^{(0)\, a}{}_{b}\right)
\left[
  2R^{(0)}_{(-)\, \mu\nu}{}^{b}{}_{a}
      -\left(kF_{[\mu|}{}^{b}-k^{-1}G^{(0)}{}_{[\mu}{}^{b}\right)
      \left(kF_{\nu]\, a}-k^{-1}G^{(0)}{}_{|\nu]\, a}\right)
    \right.
    \\
    & \\
    & \hspace{.5cm}
    \left.
      \left.
-\tfrac{1}{2}\left(kF^{b}{}_{a}+k^{-1}G^{(0)\, b}{}_{a}\right)kF_{\mu\nu} \right]
    -2\partial^{a}k\left[2\mathcal{D}^{(0)}_{(-)\, [\mu}\left(kF_{\nu]\, a}
      -k^{-1}G^{(0)}{}_{|\nu]\, a}\right)
      -\partial_{a}k F_{\mu\nu} \right]
\right\}
   \, .
      \end{aligned}
\end{equation}

\noindent
Since the right-hand side has to be a gauge-invariant combination, it is
natural to define the first-order in $\alpha'$ winding vector and its field
strength by

\begin{subequations}
  \begin{align}
    B^{(1)}{}_{\mu}
    & \equiv
      \hat{B}_{\mu\, \underline{z}}
      -\frac{\alpha'}{4}k\left(\varphi_{A}A^{A}{}_{\mu}
      +\varphi^{\hat{a}}{}_{\hat{b}}A^{\hat{b}}{}_{\hat{a}\, \mu}\right)
      \nonumber \\
    & \nonumber \\
    & =
      \hat{B}_{\mu\, \underline{z}}
      -\frac{\alpha'}{4}\left[\hat{A}^{A}{}_{\mu}\hat{A}_{A\, \underline{z}}
      +\hat{\Omega}^{(0)}_{(-)\, \mu}{}^{\hat{a}}{}_{\hat{b}}
      \hat{\Omega}^{(0)}_{(-)\, \underline{z}}{}^{\hat{b}}{}_{\hat{a}}
      \right.
      \nonumber \\
    & \nonumber \\
    & \hspace{.5cm}
      \left.
      -A_{\mu}\left(\hat{A}^{A}{}_{\underline{z}}\hat{A}_{A\, \underline{z}}
      +\hat{\Omega}^{(0)}_{(-)\, \underline{z}}{}^{\hat{a}}{}_{\hat{b}}
      \hat{\Omega}^{(0)}_{(-)\, \underline{z}}{}^{\hat{b}}{}_{\hat{a}}\right)\right]\, ,
    \\
    & \nonumber \\
    G^{(1)}{}_{\mu\nu}
    & \equiv
      2\partial_{[\mu}B^{(1)}{}_{\nu]}\, .
  \end{align}
\end{subequations}

\noindent
Furthermore, it is also natural to define the combinations

\begin{equation}
    K^{(\pm)}{}_{\mu\nu}
    \equiv
    kF_{\mu\nu}\pm k^{-1}G^{(0)}{}_{\mu\nu}\, .
\end{equation}

\noindent
$K^{(+)}{}_{\mu\nu}$ is invariant under the zeroth-order T~duality
transformations Eq.~(\ref{eq:0thorderTduality}) while $K^{(-)}{}_{\mu\nu}$
gets a minus sign under the same transformations. With this notation, we can
finally write

\begin{equation}
  \begin{aligned}
  \hat{H}^{(1)}{}_{abz}
  & =
  k^{-1}G^{(1)}{}_{ab} 
  +\frac{\alpha'}{4}
\left\{
  2\varphi_{A}F^{A}{}_{ab}
  +\left[\varphi^{2}-\tfrac{1}{4}K^{(+)\, 2}
    +2(\partial\log{k})^{2}\right]kF_{ab} 
\right.
\\
& \\
& \hspace{.5cm}
 \left.
+R^{(0)}_{(-)\, ab}{}^{cd}K^{(+)}{}_{cd}
      -\tfrac{1}{2}K^{(-)}{}_{a}{}^{c}K^{(-)}{}_{b}{}^{d}K^{(+)}{}_{cd}
    -4\mathcal{D}^{(0)}_{(-)\, [a}K^{(-)}{}_{b]\, c}\partial^{c}\log{k}
   \right\}\, .
      \end{aligned}
\end{equation}

This term contributes as
$-\tfrac{1}{4}\hat{H}^{(1)}{}_{abz} \hat{H}^{(1)\, ab}{}_{z}$, which, at first
order in $\alpha'$ gives

\begin{equation}
  \begin{aligned}
    -\tfrac{1}{4}\hat{H}^{(1)}{}_{abz} \hat{H}^{(1)\,  ab}{}_{z}
    & =
    -\tfrac{1}{4}  k^{-2}G^{(1)\, 2}
    \\
    & \\
    & \hspace{.5cm}
  -\frac{\alpha'}{8}
\left\{
2  \varphi_{A}F^{A}\cdot k^{-1}G^{(0)}
  +\left[\varphi^{2}-\tfrac{1}{4}K^{(+)\,
      2}+2(\partial\log{k})^{2}\right]F\cdot G^{(0)} 
\right.
\\
& \\
& \hspace{.5cm}
+ k^{-1} R^{(0)}_{(-)\, ab}{}^{cd}K^{(+)}{}_{cd}G^{(0)\, ab}
      -\tfrac{1}{2}k^{-1}G^{(0)\, ab}K^{(-)}{}_{a}{}^{c}K^{(-)}{}_{b}{}^{d}K^{(+)}{}_{cd}
      \\
      & \\
      & \hspace{.5cm}
      \left.
        -4k^{-1}G^{(0)\, ab}\mathcal{D}^{(0)}_{(-)\, [a}K^{(-)}{}_{b]\, c}
        \partial^{c}\log{k}
   \right\}\, .
      \end{aligned}
\end{equation}

Let us now move to the gauge-invariant combination $\hat{H}^{(1)}{}_{abc}$,
that we will identify with the 9-dimensional Kalb-Ramond 3-form field
strength. Using the zeroth-order result, we get

\begin{equation}
  \hat{H}^{(1)}{}_{abc}
  =
  H^{(0)}{}_{abc}
  +\frac{\alpha'}{4}\left(\hat{\omega}^{\rm YM}{}_{abc}
    +\hat{\omega}^{\rm L\, (0)}{}_{abc}\right)\, .
\end{equation}

\noindent
Using Eqs.~(\ref{eq:YMreduction}) it is almost immediately seen that 

\begin{equation}
  \hat{\omega}^{\rm YM}{}_{abc}
  =
  \omega^{\rm YM}{}_{abc} +3 k\varphi_{A}F_{[ab}A^{A}{}_{c]}\, .
\end{equation}

\noindent
Half of the last term has to be integrated by parts, and the final result is

\begin{equation}
  \hat{\omega}^{\rm YM}{}_{abc}
  =
  \omega^{\rm YM}{}_{abc}
  +3e_{a}{}^{\mu}e_{b}{}^{\nu}e_{c}{}^{\rho}
  \left[  \partial_{[\mu}\left(kA_{\nu|}\varphi_{A}A^{A}{}_{|\rho]}\right)
    +A_{[\mu}\partial_{\nu|}\left(k\varphi_{A}A^{A}{}_{|\rho]}\right)
  +\tfrac{3}{2}k\varphi_{A}A^{A}{}_{[\mu}F_{\nu\rho]}\right]\, .
\end{equation}

\noindent
The second term in the above expression, a total derivative, will combine with
$\hat{B}_{\mu\nu}$ (and terms coming from
$\hat{\omega}^{\rm L\, (0)}{}_{abc}$) to give $B^{(1)}{}_{\mu\nu}$ and the
third term, as we know, combines with $\hat{B}_{\mu\underline{z}}$ (and terms
coming from $\hat{\omega}^{\rm L\, (0)}{}_{abc}$) to give $B^{(1)}{}_{\mu}$.

The above result can be applied to $\hat{\omega}^{\rm L\, (0)}{}_{abc}$, using
the definitions Eq.~(\ref{eq:lorentzgaugefields}). We get

\begin{equation}
  \begin{aligned}
    \hat{\omega}^{\rm L\, (0)}{}_{abc}
    & =
    \omega^{\rm L\, (0)}{}_{abc}     +3e_{a}{}^{\mu}e_{b}{}^{\nu}e_{c}{}^{\rho} \left[
      \mathcal{D}^{(0)}_{(-)\, [\mu}K^{(-)}{}_{\nu}{}^{e}K^{(-)}{}_{\rho]\, e}
      +\partial_{[\mu} \left(
        kA_{\nu|}\varphi^{\hat{e}}{}_{\hat{f}}A^{\hat{f}}{}_{\hat{e}\, |\rho]}
      \right)
\right.
    \\
    & \\
    & \hspace{.5cm}
\left.
    +A_{[\mu}\partial_{\nu|}
      \left(k\varphi^{\hat{e}}{}_{\hat{f}}A^{\hat{f}}{}_{\hat{e}\,
          |\rho]}\right)
      +\tfrac{1}{2}k\varphi^{\hat{e}}{}_{\hat{f}}A^{\hat{f}}{}_{\hat{e}\,[\mu}F_{\nu\rho]}\right]\, .
  \end{aligned}
\end{equation}

Defining

\begin{equation}
  B^{(1)}{}_{\mu\nu}
  \equiv
  \hat{B}_{\mu\nu}
  +A_{[\mu}
  \left[
    \hat{B}_{\nu]\, \underline{z}}
  +\frac{\alpha'}{4}
 k \left(
    \hat{A}^{A}{}_{|\nu]}\hat{A}_{A\, \underline{z}}
      +\hat{\Omega}^{(0)}_{(-)\, |\nu]}{}^{\hat{a}}{}_{\hat{b}}
      \hat{\Omega}^{(0)}_{(-)\, \underline{z}}{}^{\hat{b}}{}_{\hat{a}}
    \right)
    \right]\, ,
\end{equation}

\noindent
we find

\begin{subequations}
  \begin{align}
  \hat{H}^{(1)}{}_{abc}
  & =
    H^{(1)}{}_{abc}\, ,
    \\
    & \nonumber \\
    \label{eq:H1}
    H^{(1)}{}_{\mu\nu\rho}
    & \equiv
      3\partial_{[\mu}B^{(1)}{}_{\nu\rho]}
      -\tfrac{3}{2}A_{[\mu}G^{(1)}{}_{\nu\rho]}
      -\tfrac{3}{2}B^{(1)}{}_{[\mu}F_{\nu\rho]}
      \nonumber \\
    & \nonumber \\
    & \hspace{.5cm}
      +\frac{\alpha'}{4}\left(\omega^{\rm YM}{}_{\mu\nu\rho}+\omega^{\rm
      L\,(0)}_{(-)\, \mu\nu\rho}
      +3\mathcal{D}^{(0)}_{(-)\, [\mu}K^{(-)}{}_{\nu}{}^{e}K^{(-)}{}_{\rho]\, e} \right)\, .
  \end{align}
\end{subequations}

Summarizing, the reduction of all the terms in the action but the last one 
gives, to $\mathcal{O}(\alpha')$,

\begin{equation}
  \begin{aligned}
    & \int dz\int d^{9}x \sqrt{|g|}e^{-2\phi} \left\{ R -4(\partial\phi)^{2}
      +\left(1+\frac{\alpha'}{4}\varphi^{2}\right)(\partial \log{k})^{2}
      +\frac{\alpha'}{4}(\mathfrak{D}\varphi)^{2} \right.
          +\frac{\alpha'}{4}\partial_{a}\log{k}\partial^{a}\varphi^{2}
    \\
    & \\
    & -\tfrac{1}{4}\left(1+\frac{\alpha'}{2}\varphi^{2}\right)k^{2}F^{2}
    +\tfrac{1}{12}H^{(1)\, 2}
    -\tfrac{1}{4} k^{-2}G^{(1)\, 2} -\frac{\alpha'}{8}\left[F_{A}\cdot F^{A}
      +2\varphi_{A}F^{A}\cdot K^{(+)} \right.
    \\
    & \\
    & +\left[\varphi^{2}-\tfrac{1}{4}K^{(+)\, 2}+2(\partial\log{k})^{2}\right]
    F\cdot G^{(0)}
    +R^{(0)}_{(-)\, ab}{}^{cd}K^{(+)}{}_{cd}k^{-1}G^{(0)\, ab}
    \\
    & \\
    & \left.  \left. -\tfrac{1}{2}k^{-1}G^{(0)\,
          ab}K^{(-)}{}_{a}{}^{c}K^{(-)}{}_{b}{}^{d}K^{(+)}{}_{cd}
        -4k^{-1}G^{(0)\, ab}\mathcal{D}^{(0)}_{(-)\, [a}K^{(-)}{}_{b]\,
          c}\partial^{c}\log{k} \right] \right\}\, .
  \end{aligned}
\end{equation}

Now, we must deal with the last term. We deal with it in the same way as we dealt
with the Yang-Mills kinetic term:

\begin{equation}
  \begin{aligned}
    \hat{R}^{(0)}_{(-)\, \hat{c}\hat{d}}{}^{\hat{a}}{}_{\hat{b}}\,
    \hat{R}^{(0)}_{(-)}{}^{\hat{c}\hat{d}\, \hat{b}}{}_{\hat{a}} & =
    \hat{F}^{\hat{a}}{}_{\hat{b}\, cd} \hat{F}^{\hat{b}}{}_{\hat{a}}{}^{cd}
    -2\hat{F}^{\hat{a}}{}_{\hat{b}\, cz}
    \hat{F}^{\hat{b}}{}_{\hat{a}}{}^{c}{}_{z}
    \\
    & \\
    & = \left(F^{\hat{a}}{}_{\hat{b}\,
        cd}+k\varphi^{\hat{a}}{}_{\hat{b}}F_{cd} \right) \left(
      F^{\hat{b}}{}_{\hat{a}}{}^{cd}+k\varphi^{\hat{b}}{}_{\hat{a}}
      F^{cd}\right)
    \\
    & \\
    & \hspace{.5cm}
    -2\left( \mathcal{D}_{c}\varphi^{\hat{a}}{}_{\hat{b}}
      +\varphi^{\hat{a}}{}_{\hat{b}}\partial_{c}\log{k} \right) \left(
      \mathcal{D}^{c}\varphi^{\hat{b}}{}_{\hat{a}}
      +\varphi^{\hat{b}}{}_{\hat{a}}\partial^{c}\log{k} \right)\, .
\end{aligned}
\end{equation}

\noindent
The Lorentz-covariant derivatives in the last line must be taken with respect
to the connection $A^{\hat{a}\hat{b}}{}_{\mu}$, which means that the $ab$
components contain contributions from $A^{az}{}_{\mu}$ etc.  Taking this fact
into account, if we split the hatted indices into unhatted indices and $z$
components, we get

\begin{equation}
\begin{aligned}
  & \left(F^{a}{}_{b\, \mu\nu}+k\varphi^{a}{}_{b}F_{\mu\nu} \right) \left(
    F^{b}{}_{a}{}^{\mu\nu}+k\varphi^{b}{}_{a} F^{\mu\nu}\right) +2
  \left(F^{az}{}_{\mu\nu}+k\varphi^{az}F_{\mu\nu}\right)
  \left(F_{a}{}^{z}{}^{\mu\nu}+k\varphi_{a}{}^{z}F^{\mu\nu}\right)
  \\
  & \\
  & -2\left( \mathcal{D}_{c}\varphi^{a}{}_{b} -A^{az}{}_{c}\varphi_{b}{}^{z}
    +A_{b}{}^{z}{}_{c}\varphi^{az} +\varphi^{a}{}_{b}\partial_{c}\log{k}
  \right) \left( \mathcal{D}^{c}\varphi^{b}{}_{a}
    -A^{bz}{}_{c}\varphi_{a}{}^{z} +A_{a}{}^{z}{}_{c}\varphi^{bz}
    +\varphi^{b}{}_{a}\partial^{c}\log{k} \right)
  \\
  & \\
  & -4\left( \mathcal{D}_{c}\varphi^{az} +A^{bz}{}_{c}\varphi^{a}{}_{b}
    +\varphi^{az}\partial_{c}\log{k} \right) \left(
    \mathcal{D}^{c}\varphi_{a}{}^{z} +A^{bz\, c}\varphi_{ab}
    +\varphi_{a}{}^{z}\partial^{c}\log{k} \right)\, ,
\end{aligned}
\end{equation}

\noindent
where, now $\mathcal{D}_{c}$ is the Lorentz-covariant derivative with respect
to the connection $A^{ab}{}_{\mu}$.  

Substituting the components
$A^{ab}{}_{\mu},A^{az}{}_{\mu},\varphi^{ab},\varphi^{az}$ by their values, we get

\begin{equation}
  \begin{aligned}
    & \left( R^{(0)}_{(-)\, \mu\nu}{}^{a}{}_{b}
      -\tfrac{1}{2}K^{(-)}{}_{[\mu}{}^{a}K^{(-)}{}_{\nu]\, b} -\tfrac{1}{2}
      K^{(+)\, a}{}_{b}kF_{\mu\nu} \right) \left( R^{(0)}_{(-)}{}^{\mu\nu\,
        b}{}_{a} -\tfrac{1}{2}K^{(-)\, \mu\, b}K^{(-)\, \nu}{}_{a}
      -\tfrac{1}{2} K^{(+)\, b}{}_{a}kF^{\mu\nu} \right)
    \\
    & \\
    & +2 \left( \mathcal{D}^{(0)}_{(-)\, [\mu}K^{(-)}{}_{\nu]\, a}
      -\partial_{a}\log{k}\, kF_{\mu\nu}\right) \left(
      \mathcal{D}^{(0)}_{(-)}{}^{[\mu|}K^{(-)\, |\nu]\, a}
      -\partial^{a}\log{k}\, kF^{\mu\nu} \right)
    \\
    & \\
    & +\tfrac{1}{2}\left( \mathcal{D}^{(0)}_{(-)}{}^{c}K^{(+)\, ab} -2K^{(-)\,
        c\, [a}\partial^{b]}\log{k} +K^{(+)\, ab}\partial^{c}\log{k} \right)
    \\
    & \\
    & \left( \mathcal{D}^{(0)}_{(-)\, c}K^{(+)}{}_{ab} -2K^{(-)}{}_{c\,
        [a}\partial_{b]}\log{k} +K^{(+)}{}_{ab}\partial_{c}\log{k} \right)
    \\
    & \\
    & -4\left( \mathcal{D}^{(0)}_{(-)}{}^{c}\partial^{a}\log{k}
      -\tfrac{1}{4}K^{(-)\, cb}K^{(+)}{}_{b}{}^{a}
      +\partial^{a}\log{k}\partial^{c}\log{k} \right)
    \\
    & \\
    &
    \left(
      \mathcal{D}^{(0)}_{(-)\, c}\partial_{a}\log{k}
      -\tfrac{1}{4}K^{(-)}{}_{c}{}^{b}K^{(+)}{}_{ba}
      +\partial_{a}\log{k}\partial_{c}\log{k} \right)\, .
  \end{aligned}
\end{equation}

\noindent
Operating, we finally get

\begin{equation}
  \begin{aligned}
\hat{R}^{(0)}_{(-)\, \hat{c}\hat{d}}{}^{\hat{a}}{}_{\hat{b}}\,
\hat{R}^{(0)}_{(-)}{}^{\hat{c}\hat{d}\, \hat{b}}{}_{\hat{a}}
& =
R^{(0)}_{(-)\,  \mu\nu}{}^{a}{}_{b} R^{(0)}_{(-)}{}^{\mu\nu\, b}{}_{a}
+R^{(0)}_{(-)\,  \mu\nu}{}^{ab}K^{(-)\, \mu}{}_{a}K^{(-)\, \nu}{}_{b}
+R^{(0)}_{(-)\,  \mu\nu}{}^{ab}K^{(+)}{}_{ab}kF^{\mu\nu}
\\
& \\
& \hspace{.5cm}
+\tfrac{1}{4}K^{(-)}{}_{[\mu|}{}^{a}K^{(-)}{}_{a}{}^{\nu}K^{(-)}{}_{|\nu]}{}^{b}
K^{(-)}{}_{b}{}^{\mu}
-\tfrac{1}{2}K^{(-)}{}_{\mu a}K^{(-)}{}_{\nu b}K^{(+)\, ab}kF^{\mu\nu}
\\
& \\
& \hspace{.5cm}
-\tfrac{1}{4}(K^{(+)})^{2}k^{2}F^{2}
+2\mathcal{D}^{(0)}_{(-)}{}^{[\mu|}K^{(-)\, |\nu]\, a}\mathcal{D}^{(0)}_{(-)\,
  [\mu|}K^{(-)}{}_{|\nu]\, a}
\\
& \\
& \hspace{.5cm}
-4\mathcal{D}^{(0)}_{(-)}{}^{\mu}K^{(-)\, \nu a}\partial_{a}\log{k} kF_{\mu\nu}
+2(\partial\log{k})^{2}k^{2}F^{2}
\\
& \\
& \hspace{.5cm}
+\tfrac{1}{2}
\mathcal{D}^{(0)}_{(-)}{}^{c}K^{(+)\, ab}\mathcal{D}^{(0)}_{(-)\, c}K^{(+)}{}_{ab}
-2 \mathcal{D}^{(0)}_{(-)}{}^{c}K^{(+)\, ab}K^{(-)}{}_{c\,
  a}\partial_{b}\log{k}
\\
& \\
& \hspace{.5cm}
+\mathcal{D}^{(0)}_{(-)}{}^{c}K^{(+)\, ab}K^{(+)}{}_{ab}\partial_{c}\log{k}
+2K^{(-)\, c\, [a}\partial^{b]}\log{k}K^{(-)}{}_{ca}\partial_{b}\log{k}
\\
& \\
& \hspace{.5cm}
%-2K^{(-)\, c\, a}\partial^{b}\log{k}K^{(+)}{}_{ab}\partial_{c}\log{k}
+\tfrac{1}{2}(K^{(+)})^{2}(\partial\log{k})^{2}
\\
& \\
& \hspace{.5cm}
-4\mathcal{D}^{(0)}_{(-)}{}^{c}\partial^{a}\log{k}
\mathcal{D}^{(0)}_{(-)\, c}\partial_{a}\log{k}
+2\mathcal{D}^{(0)}_{(-)}{}^{c}\partial^{a}\log{k}K^{(-)}{}_{c}{}^{b}K^{(+)}{}_{ba}
\\
& \\
& \hspace{.5cm}
% +2K^{(-)\, cb}K^{(+)}{}_{b}{}^{a}\partial_{c}\log{k}\partial_{a}\log{k}
-\tfrac{1}{4}K^{(-)\, a}{}_{b}K^{(+)\, b}{}_{c}K^{(-)\, c}{}_{d}K^{(+)\, d}{}_{a}
\\
& \\
& \hspace{.5cm}
-8\mathcal{D}^{(0)}_{(-)}{}^{c}\partial^{a}\log{k}
\partial_{a}\log{k}\partial_{c}\log{k}
-4((\partial\log{k})^{2})^{2}\, .
\end{aligned}
\end{equation}

With all these terms, the action takes the form

\begin{equation}
  \label{eq:9dimensionalheterotic}
  \begin{aligned}
    S    & =
      \frac{g_{s}^{2}(2\pi\ell_{s})}{16\pi G_{N}^{(10)}}
\int d^{9}x \sqrt{|g|}e^{-2\phi} \left\{ R -4(\partial\phi)^{2}
      +\frac{\alpha'}{4}(\mathfrak{D}\varphi)^{2}
      -\partial_{a}k^{-1}\partial^{a}k^{(1)} \right.
    \\
    & \\
    & \hspace{.5cm}
    -\tfrac{1}{4}k^{(1)\, 2}F^{2} -\tfrac{1}{4} k^{-2}G^{(1)\, 2}
+\tfrac{1}{2}(1-k^{(1)}k^{-1})F\cdot G^{(1)}
    +\tfrac{1}{12}H^{(1)\,  2}
    \\
    & \\
    & \hspace{.5cm}
    -\frac{\alpha'}{8}\left[ F_{A}\cdot
      F^{A}+R^{(0)}_{(-)}{}^{a}{}_{b}\cdot R^{(0)}_{(-)}{}^{b}{}_{a}
      +R^{(0)}_{(-)\, ab}{}^{cd}\left(K^{(-)\, a}{}_{c}K^{(-)\, b}{}_{d}
        +K^{(+)\, ab}K^{(+)}{}_{cd}\right) \right.
    \\
    & \\
    & \hspace{.5cm}
    % +2\left[\varphi^{2}-\tfrac{1}{4}K^{(+)\, 2}+2(\partial\log{k})^{2}\right]F\cdot G^{(0)}
    +2\varphi_{A}F^{A}\cdot
    K^{(+)}
    \\
    & \\
    & \hspace{.5cm}
    -\tfrac{3}{4}K^{(+)\, a}{}_{b}K^{(-)\, b}{}_{c}K^{(+)\, c}{}_{d}K^{(-)\,d}{}_{a}
    +\tfrac{1}{8}K^{(-)\, a}{}_{b}K^{(-)\, b}{}_{c}K^{(-)\,c}{}_{d}K^{(-)\,d}{}_{a}
    -\tfrac{1}{8}\left(K^{(-)}\cdot K^{(-)}\right)^{2}
    \\
    & \\
    & \hspace{.5cm}
    -4K^{(+)\, ab}\mathcal{D}^{(0)}_{(-)\,
      a}K^{(-)}{}_{bc}\partial^{c}\log{k} -2K^{(-)\, ab}
    \mathcal{D}^{(0)}_{(-)\, a}K^{(+)}{}_{bc}\partial^{c}\log{k}
    \\
    & \\
    & \hspace{.5cm}
    +2\mathcal{D}^{(0)}_{(-)}{}^{[a|}K^{(-)\, |b]\, c}
    \mathcal{D}^{(0)}_{(-)\, [a|}K^{(-)}{}_{|b]\, c}
    +\tfrac{1}{2}\mathcal{D}^{(0)}_{(-)}{}^{c}K^{(+)\, ab}
    \mathcal{D}^{(0)}_{(-)\, c}K^{(+)}{}_{ab}
    \\
    & \\
    & \hspace{.5cm}
    -4\mathcal{D}^{(0)}_{(-)}{}^{c}\partial^{a}\log{k}
    \mathcal{D}^{(0)}_{(-)\, c}\partial_{a}\log{k}
    +2 K^{(-)\, ac}K^{(+)}{}_{c}{}^{b}
    \mathcal{D}^{(0)}_{(-)\, a}\partial_{b}\log{k}
    \\
    & \\
    & \hspace{.5cm}
    \left.  \left.  +2K^{(-)\, c\,
          [a}\partial^{b]}\log{k}K^{(-)}{}_{ca}\partial_{b}\log{k}  \right]
    \right\}\, ,
  \end{aligned}
\end{equation}

\noindent\
where we have defined

\begin{equation}
k^{(1)} \equiv k\left[1+\frac{\alpha'}{4}\left(\varphi^{2}-\tfrac{1}{4}K^{(+)\, 2}
      +2(\partial\log{k})^{2}\right)\right]\, , 
\end{equation}

\noindent
and we have added some $\mathcal{O}(\alpha'{}^{2})$ terms in order to obtain
nicer or simpler expressions.

%%%%%%%%%%%%%%%%%%%%%%%%%%%%%%%%%%%%%%%%%%%%%%%%%%%%%%%%%%%%%%%%%%%%%% 
%%%%%%%%%%%%%%%%%%%%%%%%%%%%%%%%%%%%%%%%%%%%%%%%%%%%%%%%%%%%%%%%%%%%%%
%%%%%%%%%%%%%%%%%%%%%%%%%%%%%%%%%%%%%%%%%%%%%%%%%%%%%%%%%%%%%%%%%%%%%%
%%%%%%%%%%%%%%%%%%%%%%%%%%%%%%%%%%%%%%%%%%%%%%%%%%%%%%%%%%%%%%%%%%%%%%
\subsection{T~duality}
\label{sec-Tduality}
%%%%%%%%%%%%%%%%%%%%%%%%%%%%%%%%%%%%%%%%%%%%%%%%%%%%%%%%%%%%%%%%%%%%%% 
%%%%%%%%%%%%%%%%%%%%%%%%%%%%%%%%%%%%%%%%%%%%%%%%%%%%%%%%%%%%%%%%%%%%%%
%%%%%%%%%%%%%%%%%%%%%%%%%%%%%%%%%%%%%%%%%%%%%%%%%%%%%%%%%%%%%%%%%%%%%%
%%%%%%%%%%%%%%%%%%%%%%%%%%%%%%%%%%%%%%%%%%%%%%%%%%%%%%%%%%%%%%%%%%%%%%

All the $\mathcal{O}(\alpha')$ terms of the reduced action
Eq.~(\ref{eq:9dimensionalheterotic}) are invariant under the zeroth-order
T~duality transformations Eqs.~(\ref{eq:0thorderTduality}), and the whole
action is invariant to $\mathcal{O}(\alpha')$ under the transformations

\begin{equation}
  \label{eq:1storderTduality}
  A_{\mu}'
  = B^{(1)}{}_{\mu}\, , 
  \hspace{1cm}
  B^{(1)}{}_{\mu}'
  = 
  A_{\mu}\, ,
  \hspace{1cm}
  k'
   = 
  1/k^{(1)}\, ,
\end{equation}

\noindent
which reduce to the zeroth-order ones in Eqs.~(\ref{eq:0thorderTduality}) when
we set $\alpha'=0$. Furthermore, observe that

\begin{equation}
  \begin{aligned}
    k^{(1)\,\prime}
    & =
    k' \left[1+\frac{\alpha'}{4}\left(\varphi^{2}-\tfrac{1}{4}K^{(+)\, 2}
        +2(\partial\log{k})^{2}\right)\right]
    \\
    & \\
    & =
    k^{(1)\,-1}\left[1+\frac{\alpha'}{4}\left(\varphi^{2}-\tfrac{1}{4}K^{(+)\,
          2} +2(\partial\log{k})^{2}\right)\right]
    \\
    & \\
    & =
    k^{-1}\left[1 +\mathcal{O}(\alpha^{\prime\, 2}) \right]\, .
  \end{aligned}
\end{equation}

Using the relation between the higher- and lower-dimensional fields, these
transformations can be expressed in terms of the higher-dimensional ones in
the form

\begin{equation}
\label{eq:buscheralphaprime}
\begin{aligned}
  \hat{g}'_{\mu\nu}
  & = 
        \hat{g}_{\mu\nu}+\frac{\hat{g}_{\underline{z}\underline{z}} \hat{\mathfrak{G}}^{(1)}{}_{\underline{z}\mu}\hat{\mathfrak{G}}^{(1)}{}_{\underline{z}\nu}}{\hat{\mathfrak{G}}^{(1)}{}_{\underline{z}\underline{z}}^{2}}
-\frac{2\hat{\mathfrak{G}}^{(1)}{}_{\underline{z}(\mu}
\hat{g}_{\nu)\underline{z}}}{\hat{\mathfrak{G}}^{(1)}{}_{\underline{z}\underline{z}}}\, ,
        \hspace{-2cm} & & \\
 & \\
  \hat{B}'_{\mu\nu}
  & = 
  \hat{B}_{\mu\nu}
  -\frac{\hat{\mathfrak{G}}^{(1)}{}_{\underline{z}[\mu} \hat{\mathfrak{G}}^{(1)}{}_{\nu]\underline{z}}}{\hat{\mathfrak{G}}^{(1)}{}_{\underline{z}\underline{z}}}\, ,
  \\
  & \\
  \hat{g}'_{\underline{z}\mu}
  & = 
-\frac{\hat{g}_{\underline{z}\mu}}{\hat{\mathfrak{G}}^{(1)}{}_{\underline{z}\underline{z}}}
+\frac{\hat{g}_{\underline{z}\underline{z}}\hat{\mathfrak{G}}^{(1)}{}_{\underline{z}\mu}}{\hat{\mathfrak{G}}^{(1)}{}_{\underline{z}\underline{z}}^{2}}\, ,
        \hspace{1cm}
  &
    \hat{B}'_{\underline{z}\mu}
    & = 
          -\frac{\hat{B}_{\underline{z}\mu}}{\hat{\mathfrak{G}}^{(1)}{}_{\underline{z}\underline{z}}}
          -\frac{\hat{\mathfrak{G}}^{(1)}{}_{\underline{z}\mu}}{\hat{\mathfrak{G}}^{(1)}{}_{\underline{z}\underline{z}}}\, ,
  \\
  &  \\
  \hat{g}'_{\underline{z}\underline{z}}
  & = 
\frac{\hat{g}_{\underline{z}\underline{z}}}{\hat{\mathfrak{G}}^{(1)}{}_{\underline{z}\underline{z}}^{2}}\, ,
  &
    e^{-2\hat{\phi}'}
    & = 
          e^{-2\hat{\phi}}|\hat{\mathfrak{G}}^{(1)}{}_{\underline{z}\underline{z}}|\, ,
  \\
  & \\
  \hat{A}'^{A}{}_{\underline{z}}
  & = 
        -\frac{\hat{A}^{A}{}_{\underline{z}}}{\hat{\mathfrak{G}}^{(1)}{}_{\underline{z}\underline{z}}}\, ,
  &
    \hat{A}'^{A}{}_{\mu}
    & = 
          \hat{A}^{A}{}_{\mu}
          -\frac{\hat{A}^{A}{}_{\underline{z}}\hat{\mathfrak{G}}^{(1)}{}_{\underline{z}\mu}}{\hat{\mathfrak{G}}^{(1)}{}_{\underline{z}\underline{z}}}\, ,
\end{aligned}
\end{equation}

\noindent
where the  tensor $\hat{\mathfrak{G}}^{(1)}{}_{\hat{\mu}\hat{\nu}}$ is defined by

\begin{equation}
  \hat{\mathfrak{G}}^{(1)}{}_{\hat{\mu}\hat{\nu}}
  \equiv
  \hat{g}_{\hat{\mu}\hat{\nu}} -\hat{B}_{\hat{\mu}\hat{\nu}}
  -\frac{\alpha'}{4} \left\{
    \hat{A}^{A}{}_{\hat{\mu}}^{A}\hat{A}_{A\, \hat{\nu}}
    +\hat{\Omega}^{(0)}_{(-)\, \hat{\mu}}{}^{\hat{a}}{}_{\hat{b}}
    \hat{\Omega}^{(0)}_{(-)\, \hat{\nu}}{}^{\hat{b}}{}_{\hat{a}} \right\}\, .
\end{equation}

These are the $\alpha'$-corrected Buscher rules first found in
Ref.~\cite{Bergshoeff:1995cg} and later rediscovered elsewhere
\cite{Serone:2005ge,Bedoya:2014pma}. 

It is well known that $\mathcal{N}=1,d=10$ supergravity
\cite{Bergshoeff:1981um,Chapline:1982ww} coupled to $n_{V}$ Abelian vector
multiplets \cite{Bergshoeff:1981um,Chapline:1982ww} and dimensionally reduced
on a T$^{n}$ has a global O$(n,n+n_{V})$ symmetry which was shown in
Ref.~\cite{Maharana:1992my} to be related to string T~duality. In the case at
hand, the YM vectors are, generically, non-Abelian, which reduces the symmetry
to just O$(n,n)$ \cite{Hohm:2014eba} or just O$(1,1)$ here. This group
consists of the discrete transformation that give rise to the Buscher rules
Eq.~(\ref{eq:1storderTduality}) and rescalings of just certain
lower-dimensional fields:

\begin{equation}
  A_{\mu}' = \lambda^{-1}A_{\mu}\, ,
  \hspace{1cm}
  B^{(1)\,\prime}{}_{\mu} = \lambda B^{(1)\,\prime}{}_{\mu}\, ,
  \hspace{1cm}
  k' = \lambda k\, .
\end{equation}

Under these rescalings $K^{\pm},H^{(1)}$ and the Lorentz curvature terms
remain invariant while

\begin{equation}
k^{(1)\,\prime} = \lambda k^{(1)}\, .
\end{equation}

It can be checked that the dimensionally-reduced action
Eq.~(\ref{eq:9dimensionalheterotic}) is invariant under these transformations
and, therefore, under the whole O$(1,1)$ group.

We observe that the kinetic term of the KK and winding vectors is the sum of
two separately O$(1,1)$-invariant terms

\begin{equation}
  -\tfrac{1}{4}
  (F_{\mu\nu}\,\,\,,\,\,\,G^{(1)}{}_{\mu\nu})
  \left(
    \begin{array}{cc}
      k^{(1)\,2} & 0 \\
      & \\
      0 &  1/k^{2} \\
    \end{array}
  \right)
  \left(
    \begin{array}{c}
      F^{\mu\nu} \\ \\ G^{(1)\, \mu\nu} \\
    \end{array}
  \right)
  + \tfrac{1}{2}(1-k^{(1)}/k)F\cdot G^{(1)}\, ,
\end{equation}

\noindent
and that the diagonal kinetic matrix transforms consistently under O$(1,1)$
transformations even though, as different to the zeroth-order case, the
kinetic matrix is not an O$(1,1)$ matrix itself. The consistency is related to
the fact that it is part of a O$(1,1+n_{V})$ matrix.

%%%%%%%%%%%%%%%%%%%%%%%%%%%%%%%%%%%%%%%%%%%%%%%%%%%%%%%%%%%%%%%%%%%%%% 
%%%%%%%%%%%%%%%%%%%%%%%%%%%%%%%%%%%%%%%%%%%%%%%%%%%%%%%%%%%%%%%%%%%%%%
%%%%%%%%%%%%%%%%%%%%%%%%%%%%%%%%%%%%%%%%%%%%%%%%%%%%%%%%%%%%%%%%%%%%%%
%%%%%%%%%%%%%%%%%%%%%%%%%%%%%%%%%%%%%%%%%%%%%%%%%%%%%%%%%%%%%%%%%%%%%%
\section{Entropy formula}
\label{sec-entropyformula}
%%%%%%%%%%%%%%%%%%%%%%%%%%%%%%%%%%%%%%%%%%%%%%%%%%%%%%%%%%%%%%%%%%%%%% 
%%%%%%%%%%%%%%%%%%%%%%%%%%%%%%%%%%%%%%%%%%%%%%%%%%%%%%%%%%%%%%%%%%%%%%
%%%%%%%%%%%%%%%%%%%%%%%%%%%%%%%%%%%%%%%%%%%%%%%%%%%%%%%%%%%%%%%%%%%%%%
%%%%%%%%%%%%%%%%%%%%%%%%%%%%%%%%%%%%%%%%%%%%%%%%%%%%%%%%%%%%%%%%%%%%%%

We can use the dimensionally reduced action we have obtained to calculate the
entropy of some $d$-dimensional heterotic string black holes using the
Iyer-Wald prescription \cite{Wald:1993nt,Iyer:1994ys}. These black holes must
be solutions of the theory defined by the action
Eq.~(\ref{eq:9dimensionalheterotic}) understood as a $d$-dimensional
action. Therefore, they must be solutions of the theory defined by the action
Eq.~(\ref{heterotic}) understood as a $(d+1)$-dimensional action\footnote{The
  constant in front of the action should now contain the volume of a
  $(10-d)$-dimensional torus instead of that of circle, that is
  \begin{equation}
    \frac{g_{s}^{2}(2\pi\ell_{s})^{10-d}}{16\pi G_{N}^{(10)}}
    =
     \frac{(g^{(d)}_{s})^{2}}{16\pi G_{N}^{(d)}}\, , 
   \end{equation}
   where $g^{(d)}_{s}$ is the $d$-dimensional string coupling constant or the
   vacuum expected value of the $d$-dimensional dilaton
   $<e^{\phi}>=e^{\phi_{\infty}}$ and $G_{N}^{(d)}$ the $d$-dimensional Newton
   constant. The relations of the 10-dimensional and $d$-dimensional ones with
   the volume of the $(10-d)$-dimensional compact space, $V_{10-d}$  is
      \begin{subequations}
     \begin{align}
       g_{s}^{2} & = V_{10-d}/(2\pi\ell_{s})^{10-d}g_{s}^{(d)\, 2}\, , \\
                 & \nonumber \\
     \label{eq:relationsbetweenconstants}
        G_{N}^{(10)} & = G_{N}^{(d)} V_{10-d}\, .
     \end{align}
   \end{subequations}
 } admitting an isometry. Since this $(d+1)$-dimensional action
 can be obtained from the 10-dimensional one by a trivial compactification on
 a $10-(d+1)$-dimensional torus, the metrics of the 10-dimensional solutions
 corresponding to the $d$-dimensional black holes are the direct products of
 non-trivial $(d+1)$-dimensional metrics and the metric of a
 $10-(d+1)$-dimensional torus. The non-extremal 4-dimensional
 Reissner-Nordstr\"om black hole of Ref.~\cite{Cano:2019ycn} or the heterotic
 version of the 5-dimensional Strominger-Vafa black hole of
 Ref.~\cite{Cano:2018qev} are two interesting examples of this kind of
 solution.

 Applying directly the Iyer-Wald prescription to the $d$-dimensional action
 Eq.~(\ref{eq:9dimensionalheterotic}) we obtain the following entropy formula
 expressed in string-frame variables:

\begin{subequations}
  \begin{align}
  \label{eq:entropyd4}
S
  & =
  -2\pi\int_\Sigma d^{d-2}x\sqrt{|h|}
  \frac{\partial \mathcal{L}}{\partial R_{abcd}}
    \epsilon_{ab}\epsilon_{cd}\, ,
    \\
    & \nonumber \\
\label{eq:entropyformulastringframe}
      \frac{\partial\mathcal{L}}{\partial R_{abcd}}
    & =
      \frac{e^{-2(\phi-\phi_{\infty})}}{16\pi G_{N}^{(d)}}
      \left\{ g^{ab,\, cd}
      -\frac{\alpha'}{8} \left[H^{(0)\, abg}
        \left(\omega_{g}{}^{cd}-H^{(0)}_{g}{}^{cd}\right)
      \right.\right.
\nonumber  \\
    & \nonumber \\
    & \hspace{.5cm}
      \left. \left.
        -2R_{(-)}^{(0)\, abcd}
      +K^{(-)\, [a|c}K^{(-)\, |b]d} +K^{(+)\, ab}K^{(+)\, cd}
      \right]
      \right\}\, ,
  \end{align}
\end{subequations}

\noindent
where $|h|$ is the absolute value of the determinant of the metric induced
over the event horizon, $g^{ab,cd}= \tfrac{1}{2}(g^{ac}g^{bd}-g^{ad}g^{bc})$,
$\epsilon^{ab}$ is the event horizon's binormal normalized so that
$\epsilon_{ab}\epsilon^{ab}=-2$ and $R_{abcd}$ is the Riemann tensor.
% For the
% sake of completeness, in Appendix~\ref{sec-entropyenisteinframe} we give this
% entropy formula in the Einstein frame.

%%%%%%%%%%%%%%%%%%%%%%%%%%%%%%%%%%%%%%%%%%%%%%%%%%%%%%%%%%%%%%%%%%%%%%
%%%%%%%%%%%%%%%%%%%%%%%%%%%%%%%%%%%%%%%%%%%%%%%%%%%%%%%%%%%%%%%%%%%%%%
%%%%%%%%%%%%%%%%%%%%%%%%%%%%%%%%%%%%%%%%%%%%%%%%%%%%%%%%%%%%%%%%%%%%%%
%%%%%%%%%%%%%%%%%%%%%%%%%%%%%%%%%%%%%%%%%%%%%%%%%%%%%%%%%%%%%%%%%%%%%%
\subsection{The Wald entropy of the $\alpha'$-corrected Strominger-Vafa black hole}
\label{sec-entropystromingervafa}
%%%%%%%%%%%%%%%%%%%%%%%%%%%%%%%%%%%%%%%%%%%%%%%%%%%%%%%%%%%%%%%%%%%%%%
%%%%%%%%%%%%%%%%%%%%%%%%%%%%%%%%%%%%%%%%%%%%%%%%%%%%%%%%%%%%%%%%%%%%%%
%%%%%%%%%%%%%%%%%%%%%%%%%%%%%%%%%%%%%%%%%%%%%%%%%%%%%%%%%%%%%%%%%%%%%%
%%%%%%%%%%%%%%%%%%%%%%%%%%%%%%%%%%%%%%%%%%%%%%%%%%%%%%%%%%%%%%%%%%%%%%

The entropy formula Eq.~(\ref{eq:entropyformulastringframe}) has been shown in
Ref.~\cite{Cano:2019ycn} to give an entropy which is related to the Hawking
temperature by the thermodynamic relation

\begin{equation}
\frac{\partial S}{\partial M}=\frac{1}{T}\, ,
\end{equation}

\noindent
for the particular case of $\alpha'$-corrected, 4-dimensional, non-extremal
Reissner-Nordstr\"om black holes. In this section we want to recalculate the
Wald entropy of the $\alpha'$-corrected Strominger-Vafa black hole. Being an
extremal black hole, we will not be able to check that the entropy obtained is
related to the temperature as above, but, instead, we will be able to compare
with other results obtained in the literature and with the microscopic
calculations.

The 5-dimensional $\alpha'$-corrected Strominger-Vafa black hole corresponds
to the 10-dimensional solution of the Heterotic Superstring effective action
\cite{Cano:2018qev,Chimento:2018kop}

\begin{subequations}
\label{eq:10dsolution}
\begin{align}
d\hat{s}^{2}
& = 
\frac{2}{\mathcal{Z}_{-}}du\left(dv-\tfrac{1}{2}\mathcal{Z}_{+}du\right)
-\mathcal{Z}_{0}(d\rho^{2}+\rho^{2}d\Omega_{(3)}^{2})-dy^{i}dy^{i}\, ,
\hspace{.5cm}
i=1,\ldots,4\, ,
\\
& \nonumber \\
\hat{H}^{(1)}
& = 
d\mathcal{Z}_{-}^{-1}\wedge du\wedge dv +\star_{4}d\mathcal{Z}_{0}\, ,
\\
& \nonumber \\
e^{-2\hat{\phi}}
& = 
e^{-2\hat{\phi}_{\infty}}
\mathcal{Z}_{-}/\mathcal{Z}_{0}\, ,
\end{align}
\end{subequations}

\noindent
where $\star_{4}$ stands for the Hodge dual in the 4-dimensional Euclidean
space with metric $d\rho^{2}+\rho^{2}d\Omega_{(3)}^{2}$, and where the
$\mathcal{Z}$ functions take the values\footnote{The Regge slope parameter
  $\alpha'$ in Refs.~\cite{Cano:2018qev,Chimento:2018kop} has been replaced by
  $\alpha'/8$ here to obtain the correct form of the action and solutions.}

\begin{subequations}
\label{eq:Zs}
\begin{align}
\mathcal{Z}_{0}
& = 
1+\frac{\tilde{q}_{0}}{\rho^{2}}
-\alpha' 
\frac{\rho^{2}+2\tilde{q}_{0}}{(\rho^{2}+\tilde{q}_{0})^{2}}
+\mathcal{O}(\alpha'^{2})\, , 
\\
& \nonumber \\
\mathcal{Z}_{-}
& =
1+\frac{\tilde{q}_{-}}{\rho^{2}}
+
\mathcal{O}(\alpha'^{2})\, ,
\\
&  \nonumber \\
\mathcal{Z}_{+}
& =
1+\frac{\tilde{q}_{+}}{\rho^{2}}
+2\alpha'\frac{\tilde{q}_{+}(\rho^{2}+\tilde{q}_{0}+\tilde{q}_{-})}
{\tilde{q}_{0}(\rho^{2}+\tilde{q}_{0})(\rho^{2}+\tilde{q}_{-})}
+\mathcal{O}(\alpha'^{2})\, .
\end{align}
\end{subequations}

Compactifying this solution in a T$^{4}$ parameterized by the coordinates
$y_{i}$ is trivial. Then, we just have to compactify the resulting
6-dimensional solution to $d=5$ using the results obtained here along the
coordinate $z\equiv u/k_{\infty}$, where $k_{\infty}$ is the asymptotic value
of the KK scalar $k$. It is helpful to rewrite the 6-dimensional solution in the form

\begin{subequations}
\label{eq:6dsolution}
\begin{align}
d\hat{s}^{2}
  & =
    \frac{1}{\mathcal{Z}_{+}\mathcal{Z}_{-}}dt^{2}
    -\mathcal{Z}_{0}(d\rho^{2}+\rho^{2}d\Omega_{(3)}^{2})
    -\frac{k_{\infty}^{2}\mathcal{Z}_{+}}{\mathcal{Z}_{-}}
    \left(dz-    \frac{1}{k_{\infty}\mathcal{Z}_{+}}dt\right)^{2}\, ,
\\
& \nonumber \\
\hat{H}^{(1)}
& = 
   d\left( -\frac{k_{\infty}}{\mathcal{Z}_{-}}dt \wedge dz\right)
                +\star_{4}d\mathcal{Z}_{0}\, ,
\\
& \nonumber \\
e^{-2\hat{\phi}}
& = 
e^{-2\hat{\phi}_{\infty}}
\mathcal{Z}_{-}/\mathcal{Z}_{0}\, ,
\end{align}
\end{subequations}

\noindent
where we have set $v=t$, to identify immediately the following 5-dimensional
fields:\footnote{We have only computed $G^{(0)}$ and not $G^{(1)}$ because of
  its complication and because it is unnecessary to do it for the calculation
  of the entropy. On the other hand, the Kalb-Ramond field is customarily
  dualized into another vector field to which the third charge $\tilde{q}_{0}$
  is associated.}

\begin{subequations}
\label{eq:5dsolution}
\begin{align}
ds^{2}
  & =
    \frac{1}{\mathcal{Z}_{+}\mathcal{Z}_{-}}dt^{2}
    -\mathcal{Z}_{0}(d\rho^{2}+\rho^{2}d\Omega_{(3)}^{2})\, ,
\\
& \nonumber \\
H^{(1)}
& = 
\star_{4}d\mathcal{Z}_{0}\, ,
\\
& \nonumber \\
  F
  & =
    d\left(-\frac{1}{k_{\infty}\mathcal{Z}_{+}}dt\right)\, ,
  \\
& \nonumber \\
  G^{(0)}
  & =
    d\left(-\frac{k_{\infty}}{\mathcal{Z}_{-}}dt\right)\, ,
  \\
& \nonumber \\
e^{-2(\phi-\phi_{\infty})}
& = 
\sqrt{\mathcal{Z}_{+}\mathcal{Z}_{-}}/\mathcal{Z}_{0}\, ,
\\
& \nonumber \\
  k/k_{\infty}
  & =
    \sqrt{\mathcal{Z}_{+}/\mathcal{Z}_{-}}\, ,
\end{align}
\end{subequations}

\noindent
and the T~duality even and odd 2-forms

\begin{equation}
  K^{\pm} = -\frac{1}{\sqrt{\mathcal{Z}_{+}\mathcal{Z}_{-}}}
  \left(
    \frac{\mathcal{Z}_{+}'}{\mathcal{Z}_{+}}
  \pm  \frac{\mathcal{Z}_{-}'}{\mathcal{Z}_{-}}
  \right) d\rho \wedge dt\, ,  
\end{equation}

\noindent
where a prime indicates derivative with respect to $\rho$.

In the Vielbein basis

\begin{equation}
  e^{0} = \frac{1}{\sqrt{\mathcal{Z}_{+}\mathcal{Z}_{-}}}dt\, ,
  \hspace{.5cm}
  e^{1} = \sqrt{\mathcal{Z}_{0}}d\rho\, ,
  \hspace{.5cm}
  e^{i} = \tfrac{1}{2}\sqrt{\mathcal{Z}_{0}}\rho \theta^{i}\, ,
\end{equation}

\noindent
where the $\theta^{i}$ are the left-invariant SU$(2)$ Maurer-Cartan 1-forms
that satisfy $d\Omega^{2}{}_{(3)}=\tfrac{1}{4}\theta^{i}\theta^{i}$, the
binormal is given by just $\epsilon^{01}=+1$ and the entropy formula in
Eqs.~(\ref{eq:entropyd4}) and (\ref{eq:entropyformulastringframe}) becomes

\begin{equation}
S
   =
        \frac{1}{4G_{N}^{(5)}}\int_{\Sigma} d^{3}xe^{-2(\phi-\phi_{\infty})}\sqrt{|h|}
    \left\{1
    +\frac{\alpha'}{4} \left[
        -2R^{0101}
      +(K^{(-)\, 01})^{2}+(K^{(+)\, 01})^{2}
      \right]
      \right\}\, .
\end{equation}

The fields in the integrand are only functions of $\rho$ and we can perform
the integral over S$^{3}$. Evaluating the zeroth-order term at $\rho=0$, where
the horizon is located, we get

\begin{equation}
\begin{aligned}
  \label{eq:entropydSVbh1}
S
    &   =
      \frac{1}{4G_{N}^{(5)}}
    \left\{A_{\mathcal{H}}
    +\alpha' \pi^{2}\lim_{\rho\rightarrow 0}
      \rho^{3}\sqrt{\mathcal{Z}_{0}\mathcal{Z}_{+}\mathcal{Z}_{-}}\left[
      -\sqrt{\frac{\mathcal{Z}_{+}\mathcal{Z}_{-}}{\mathcal{Z}_{0}}}
      \left[\frac{1}{\sqrt{\mathcal{Z}_{0}}}
          \left(\frac{1}{\sqrt{\mathcal{Z}_{+}\mathcal{Z}_{-}}}\right)'
      \right]'
    \right.    \right.
    \\
    & \\
    & \hspace{.5cm}
    \left.    \left.
      +\frac{1}{\mathcal{Z}_{0}}
        \left(\frac{\mathcal{Z}_{+}'}{\mathcal{Z}_{+}}\right)^{2}
        +\frac{1}{\mathcal{Z}_{0}}
        \left(\frac{\mathcal{Z}_{-}'}{\mathcal{Z}_{-}}\right)^{2}
      \right]
      \right\}\, ,
    \end{aligned}
\end{equation}

\noindent
where $A_{\mathcal{H}}$, the area of the horizon, is given by

\begin{equation}
  A_{\mathcal{H}}
  =
  2\pi^{2}\lim_{\rho\rightarrow 0}
  \rho^{3}\sqrt{\mathcal{Z}_{0}\mathcal{Z}_{+}\mathcal{Z}_{-}}
  = 2\pi^{2}\sqrt{\tilde{q}_{0}\tilde{q}_{+}\tilde{q}_{-}}\, .
\end{equation}

\noindent
Finally, we arrive at

\begin{equation}
  \label{eq:entropydSVbh2}
  S
  =
    \frac{A_{\mathcal{H}}}{4G_{N}^{(5)}}
    \left\{ 1 +\frac{2\alpha'}{\tilde{q}_{0}}\right\}\, .
\end{equation}

In order to compare this result with the microscopic entropy in
Ref.~\cite{Kraus:2005zm}, we have to express the charges
$\tilde{q}_{+},\tilde{q}_{-},\tilde{q}_{0}$ in terms of the asymptotic
charges\footnote{See Refs.~\cite{Cano:2018qev,Faedo:2019xii}, specially
  Eqs.~(2.18),(2.20),(2.21) of the later.}. First, we have to take into
account the relation between $\tilde{q}_{+},\tilde{q}_{-},\tilde{q}_{0}$ and
the numbers of fundamental strings $n$, momentum $w$ and S5-branes $N$

\begin{equation}
  \tilde{q}_{+}
  =
  \frac{\alpha^{\prime\, 2}g_{s}^{2}n}{R_{z}^{2}}\, ,
  \hspace{.5cm}
  \tilde{q}_{-}
  =
  \alpha' g_{s}^{2}w\, ,
  \hspace{.5cm}
  \tilde{q}_{0}
  =
  \alpha' N\, .
\end{equation}

\noindent
Second, 10-dimensional Newton constant $G_{N}^{(10)}$ is given in terms of the
Regge slope parameter $\alpha'=\ell_{s}^{2}$ and the 10-dimensional string
coupling constant $g_{s}$ by

\begin{equation}
  \label{eq:d10newtonconstant}
G_{N}^{(10)}=8\pi^{6}g_{s}^{2} \alpha^{\prime\, 4}\, .
\end{equation}

This and Eq.~(\ref{eq:relationsbetweenconstants}) allow us to rewrite the 
entropy Eq.~(\ref{eq:entropydSVbh2}) in the form

\begin{equation}
  \label{eq:entropydSVbh3}
  S
  =
  2\pi \sqrt{nwN}\left(1+\frac{2}{N}\right)\, .
\end{equation}

Finally, in terms of the asymptotic charges $Q_{+},Q_{-},Q_{0}$, which are
related to the numbers of branes by

\begin{equation}
  Q_{+} = n\left(1+\frac{2}{N}\right)\,
  \hspace{.5cm}
  Q_{-}
  = w\, ,
  \hspace{.5cm}
  Q_{0}
  =
  N-1\, ,
\end{equation}

\noindent
the entropy takes the final form that can be compared with the microscopic
formula

\begin{equation}
  \label{eq:entropydSVbh4}
  S
  =
  2\pi \sqrt{Q_{+}Q_{-}(Q_{0}+3)}\, .
\end{equation}

%%%%%%%%%%%%%%%%%%%%%%%%%%%%%%%%%%%%%%%%%%%%%%%%%%%%%%%%%%%%%%%%%%%%%%
%%%%%%%%%%%%%%%%%%%%%%%%%%%%%%%%%%%%%%%%%%%%%%%%%%%%%%%%%%%%%%%%%%%%%%
%%%%%%%%%%%%%%%%%%%%%%%%%%%%%%%%%%%%%%%%%%%%%%%%%%%%%%%%%%%%%%%%%%%%%%
%%%%%%%%%%%%%%%%%%%%%%%%%%%%%%%%%%%%%%%%%%%%%%%%%%%%%%%%%%%%%%%%%%%%%%
\section{Discussion}
\label{sec-discussion}
%%%%%%%%%%%%%%%%%%%%%%%%%%%%%%%%%%%%%%%%%%%%%%%%%%%%%%%%%%%%%%%%%%%%%%
%%%%%%%%%%%%%%%%%%%%%%%%%%%%%%%%%%%%%%%%%%%%%%%%%%%%%%%%%%%%%%%%%%%%%%
%%%%%%%%%%%%%%%%%%%%%%%%%%%%%%%%%%%%%%%%%%%%%%%%%%%%%%%%%%%%%%%%%%%%%%
%%%%%%%%%%%%%%%%%%%%%%%%%%%%%%%%%%%%%%%%%%%%%%%%%%%%%%%%%%%%%%%%%%%%%%

In this paper we have performed the complete dimensional reduction of the
Heterotic Superstring effective action to first order in $\alpha'$ using the
formulation based on the supersymmetry completion of the Lorentz Chern-Simons
terms that occur in the Kalb-Ramond field strength
\cite{Bergshoeff:1988nn,Bergshoeff:1989de}. We have found a $\mathbb{Z}_{2}$
transformation of the dimensionally-reduced action that leaves it invariant
and that is an $\mathcal{O}(\alpha')$ generalization of the standard
transformations that interchange KK and winding vectors and invert the KK
scalar. In 10-dimensional variables (the components of the 10-dimensional
fields) these transformations are nothing but the $\alpha'$-corrected Buscher
rules of the Heterotic Superstring theory, first found in
\cite{Bergshoeff:1995cg}.

Then, we used the dimensionally-reduced action to find, following the
Iyer-Wald prescription \cite{Wald:1993nt,Iyer:1994ys} an entropy formula for
stringy black holes that can be obtained from a 10-dimensional solution by a
single non-trivial compactification on a circle, supplemented by a trivial
compactification on a torus. This formula was successfully applied to a
non-extremal 4-dimensional Reissner-Nordstr\"om black hole in
Ref.~\cite{Cano:2019ycn} and, in this paper, we have applied it to the
$\alpha'$-corrected heterotic version of the Strominger-Vafa black hole of
Ref.~\cite{Cano:2018qev} obtaining an entropy formula that matches the
microscopic result obtained in \cite{Kraus:2005zm} once the relations between
integration constants and asymptotic brane charges have been correctly taken
into account. As explained in Ref.~\cite{Faedo:2019xii}, the result obtained
in Ref.~\cite{Cano:2018qev} misses a factor of 2 that we recover here.

It would be desirable to obtain an entropy formula that could be applied to
more general black holes, in particular, to the 4-dimensional 4-charge
ones. This requires a more general toroidal compactification of the action
along the lines of the one recently obtained in Ref.~\cite{Eloy:2020dko},
including YM fields and closer to the Bergshoeff-de Roo formulation
\cite{kn:TOM2}. It is also necessary to prove the first law of black hole
mechanics for the Heterotic Superstring effective action in order to make sure
that the Iyer-Wald prescription can be unambiguously applied as we have done
here or that the entropy is just given by the integral of the Noether charge
as assumed in Ref.~\cite{Edelstein:2019wzg}, since, in presence of matter
fields, some terms of the total Noether charge can be related to other
terms in the first law \cite{Compere:2007vx}.

Finally, let us comment on the invariance of the black-hole temperature and
entropy under $\alpha'$-corrected T~duality, which we already mentioned in
footnote~\ref{foot:esa} in the Introduction, which was discussed in
Refs.~\cite{Horowitz:1993wt,Edelstein:2018ewc,Edelstein:2019wzg}. T~duality
manifests itself in the non-compact dimensions in which black-hole solutions
appear as such as a symmetry that only acts on some lower-dimensional vector
fields (KK and winding vector fields) and on some scalars (KK scalars and
scalars originating in the Kalb-Ramond 2-form, if one compactifies more than
one dimension). In particular, the lower-dimensional dilaton, the string
metric and the lower-dimensional Kalb-Ramond field are
T~duality-invariant.\footnote{More precisely, in toroidal compactifications,
  the lower-dimensional Kalb-Ramond 2-form is only invariant up to a
  compensating O$(n)$ gauge transformation \cite{kn:TOM2}, but its field
  strength is exactly invariant.} Since the surface gravity is computed
directly on the dimensionally-reduced metric, its invariance implies
immediately the invariance of the Hawking temperature. The invariance of the
lower-dimensional action and the invariance of the Riemann tensor (which
follows that of the metric) automatically imply the invariance of the Wald
entropy formula. In our case, the invariance of the entropy formula in
Eqs.~(\ref{eq:entropyd4}),(\ref{eq:entropyformulastringframe}) is manifest.

%%%%%%%%%%%%%%%%%%%%%%%%%%%%%%%%%%%%%%%%%%%%%%%%%%%%%%%%%%%%%%%%%%%%%%
%%%%%%%%%%%%%%%%%%%%%%%%%%%%%%%%%%%%%%%%%%%%%%%%%%%%%%%%%%%%%%%%%%%%%%
%%%%%%%%%%%%%%%%%%%%%%%%%%%%%%%%%%%%%%%%%%%%%%%%%%%%%%%%%%%%%%%%%%%%%%
%%%%%%%%%%%%%%%%%%%%%%%%%%%%%%%%%%%%%%%%%%%%%%%%%%%%%%%%%%%%%%%%%%%%%%
\section*{Acknowledgments}
%%%%%%%%%%%%%%%%%%%%%%%%%%%%%%%%%%%%%%%%%%%%%%b%%%%%%%%%%%%%%%%%%%%%%%%
%%%%%%%%%%%%%%%%%%%%%%%%%%%%%%%%%%%%%%%%%%%%%%%%%%%%%%%%%%%%%%%%%%%%%%
%%%%%%%%%%%%%%%%%%%%%%%%%%%%%%%%%%%%%%%%%%%%%%%%%%%%%%%%%%%%%%%%%%%%%%
%%%%%%%%%%%%%%%%%%%%%%%%%%%%%%%%%%%%%%%%%%%%%%%%%%%%%%%%%%%%%%%%%%%%%%

TO would like to thank Pedro F.~Ram\'{\i}rez and JJ.~Fern\'andez-Melgarejo for
many useful conversations and Pablo A.~Cano for useful comments on the
manuscript.  This work has been supported in part by the MCIU, AEI, FEDER (UE)
grant PGC2018-095205-B-I00 and by the Spanish Research Agency (Agencia Estatal
de Investigaci\'on) through the grant IFT Centro de Excelencia Severo Ochoa
SEV-2016-0597. The work of ZE has also received funding from ``la Caixa''
Foundation (ID 100010434), under the agreement LCF/BQ/DI18/11660042.  TO
wishes to thank M.M.~Fern\'andez for her permanent support.

%%%%%%%%%%%%%%%%%%%%%%%%%%%%%%%%%%%%%%%%%%%%%%%%%%%%%%%%%%%%%%%%%%%%%% 
%%%%%%%%%%%%%%%%%%%%%%%%%%%%%%%%%%%%%%%%%%%%%%%%%%%%%%%%%%%%%%%%%%%%%%
%%%%%%%%%%%%%%%%%%%%%%%%%%%%%%%%%%%%%%%%%%%%%%%%%%%%%%%%%%%%%%%%%%%%%%
%%%%%%%%%%%%%%%%%%%%%%%%%%%%%%%%%%%%%%%%%%%%%%%%%%%%%%%%%%%%%%%%%%%%%%
%%%%%%%%%%%%%%%%%%%%%%%%%%%%%%%%%%%%%%%%%%%%%%%%%%%%%%%%%%%%%%%%%%%%%%
%%%%%%%%%%%%%%%%%%%%%%%%%%%%%%%%%%%%%%%%%%%%%%%%%%%%%%%%%%%%%%%%%%%%%%
\appendix
%%%%%%%%%%%%%%%%%%%%%%%%%%%%%%%%%%%%%%%%%%%%%%%%%%%%%%%%%%%%%%%%%%%%%%
%%%%%%%%%%%%%%%%%%%%%%%%%%%%%%%%%%%%%%%%%%%%%%%%%%%%%%%%%%%%%%%%%%%%%%
%%%%%%%%%%%%%%%%%%%%%%%%%%%%%%%%%%%%%%%%%%%%%%%%%%%%%%%%%%%%%%%%%%%%%%
%%%%%%%%%%%%%%%%%%%%%%%%%%%%%%%%%%%%%%%%%%%%%%%%%%%%%%%%%%%%%%%%%%%%%%

%%%%%%%%%%%%%%%%%%%%%%%%%%%%%%%%%%%%%%%%%%%%%%%%%%%%%%%%%%%%%%%%%%%%%%
%%%%%%%%%%%%%%%%%%%%%%%%%%%%%%%%%%%%%%%%%%%%%%%%%%%%%%%%%%%%%%%%%%%%%%
%%%%%%%%%%%%%%%%%%%%%%%%%%%%%%%%%%%%%%%%%%%%%%%%%%%%%%%%%%%%%%%%%%%%%%
%%%%%%%%%%%%%%%%%%%%%%%%%%%%%%%%%%%%%%%%%%%%%%%%%%%%%%%%%%%%%%%%%%%%%%
\section{Relation between 10- and 9-dimensional fields at zeroth order in $\alpha'$}
\label{sec-10versus9atorder0}
%%%%%%%%%%%%%%%%%%%%%%%%%%%%%%%%%%%%%%%%%%%%%%%%%%%%%%%%%%%%%%%%%%%%%% 
%%%%%%%%%%%%%%%%%%%%%%%%%%%%%%%%%%%%%%%%%%%%%%%%%%%%%%%%%%%%%%%%%%%%%%
%%%%%%%%%%%%%%%%%%%%%%%%%%%%%%%%%%%%%%%%%%%%%%%%%%%%%%%%%%%%%%%%%%%%%%
%%%%%%%%%%%%%%%%%%%%%%%%%%%%%%%%%%%%%%%%%%%%%%%%%%%%%%%%%%%%%%%%%%%%%%

At zeroth order in $\alpha'$, the 10-dimensional fields can be expressed in
terms of the 9-dimensional ones as follows:

\begin{equation}
  \begin{aligned}
    \hat{g}_{\mu\nu}
    & =
    g_{\mu\nu} -k^{2}A_{\mu}A_{\nu}\, , 
    \\
    & \\
    \hat{g}_{\mu \underline{z}}
    & =
    -k^{2}A_{\mu}\, ,
    \\
    & \\
    \hat{g}_{\underline{z}\underline{z}}
    & =
    -k^{2}\, ,
    \\
    & \\
    \hat{B}_{\mu\nu}
    & =
    B^{(0)}{}_{\mu\nu} -A_{[\mu}B^{(0)}{}_{\nu]}\, ,
    \\
    & \\
    \hat{B}_{\mu\underline{z}}
    & =
    B^{(0)}{}_{\mu}\, ,
    \\
     & \\
    \hat{\phi}
    & =
    \phi+\tfrac{1}{2}\log{k}\, .
  \end{aligned}
\end{equation}

The inverse relations are

\begin{equation}
  \begin{aligned}
    g_{\mu\nu}
    & =
    \hat{g}_{\mu\nu}-\hat{g}_{\underline{z}\mu}
    \hat{g}_{\underline{z}\nu}/\hat{g}_{\underline{z}\underline{z}}\, ,
    \\
    & \\
    A_{\mu}
    & =
    \hat{g}_{\mu \underline{z}}/\hat{g}_{\underline{z}\underline{z}}\, ,
    \\
    & \\
    k
    & =
    |\hat{g}_{\underline{z}\underline{z}}|^{1/2}\, ,
    \\
    & \\
    B^{(0)}{}_{\mu\nu}
    & =
    \hat{B}_{\mu\nu}
    +\hat{g}_{\underline{z}[\mu}\hat{B}_{\nu]\underline{z}}/\hat{g}_{\underline{z}\underline{z}}\, ,
    \\
    & \\
    B^{(0)}{}_{\mu}
    & =
    \hat{B}_{\mu\underline{z}} \, ,
    \\
    & \\
    \phi
    & =
    \hat{\phi}-\tfrac{1}{4}\log{(-\hat{g}_{\underline{z}\underline{z}})}\, .
  \end{aligned}
\end{equation}

%%%%%%%%%%%%%%%%%%%%%%%%%%%%%%%%%%%%%%%%%%%%%%%%%%%%%%%%%%%%%%%%%%%%%%
%%%%%%%%%%%%%%%%%%%%%%%%%%%%%%%%%%%%%%%%%%%%%%%%%%%%%%%%%%%%%%%%%%%%%%
%%%%%%%%%%%%%%%%%%%%%%%%%%%%%%%%%%%%%%%%%%%%%%%%%%%%%%%%%%%%%%%%%%%%%%
%%%%%%%%%%%%%%%%%%%%%%%%%%%%%%%%%%%%%%%%%%%%%%%%%%%%%%%%%%%%%%%%%%%%%%
\section{Relation between 10- and 9-dimensional fields at $\mathcal{O}(\alpha')$}
\label{app-dictionaryfirstorder}
%%%%%%%%%%%%%%%%%%%%%%%%%%%%%%%%%%%%%%%%%%%%%%%%%%%%%%%%%%%%%%%%%%%%%% 
%%%%%%%%%%%%%%%%%%%%%%%%%%%%%%%%%%%%%%%%%%%%%%%%%%%%%%%%%%%%%%%%%%%%%%
%%%%%%%%%%%%%%%%%%%%%%%%%%%%%%%%%%%%%%%%%%%%%%%%%%%%%%%%%%%%%%%%%%%%%%
%%%%%%%%%%%%%%%%%%%%%%%%%%%%%%%%%%%%%%%%%%%%%%%%%%%%%%%%%%%%%%%%%%%%%%

At first order in $\alpha'$, the 10-dimensional fields can be expressed in
terms of the 9-dimensional ones as follows:

\begin{equation}
  \begin{aligned}
    \hat{g}_{\mu\nu} & =
    g_{\mu\nu} -k^{2}A_{\mu}A_{\nu}\, ,
    \\
    & \\
    \hat{g}_{\mu \underline{z}}  & =
    -k^{2}A_{\mu}\, ,
    \\
        & \\
    \hat{g}_{\underline{z}\underline{z}}
    & =
    -k^{2}\, ,
    \\
        & \\
         \hat{B}_{\mu\nu} & =
     B^{(1)}{}_{\mu\nu}
  -A_{[\mu}
  \left[
    B^{(1)}{}_{\nu]}
    +\frac{\alpha'}{2}
    k\left(\varphi_{A}A^{A}{}_{|\nu]}
        +\tfrac{1}{2}\Omega^{(0)}_{(-)\, |\nu]}{}^{ab}K^{(+)}{}_{ab}
      -K^{(-)}{}_{|\nu]}{}^{a}\partial_{a}\log{k}\right)
    \right]\, ,
    \\
        & \\
     \hat{B}_{\mu\underline{z}}
     & =
     B^{(1)}{}_{\mu}
      +\frac{\alpha'}{4}k\left(\varphi_{A}A^{A}{}_{\mu}
        +\tfrac{1}{2}\Omega^{(0)}_{(-)\, \mu}{}^{ab}K^{(+)}{}_{ab}
        -K^{(-)}{}_{\mu}{}^{a}\partial_{a}\log{k}\right)\, ,
      \\
          & \\
    \hat{\phi}
    & =
    \phi+\tfrac{1}{2}\log{k}\, ,
    \\
    & \\
    \hat{A}^{A}{}_{\mu}
    & =
    A^{A}{}_{\mu} +k\varphi^{A}A_{\mu}\, ,
    \\
    & \\
    \hat{A}^{A}{}_{\underline{z}}
    & =
    k\varphi^{A}\, .
  \end{aligned}
\end{equation}

The inverse relations are

\begin{equation}
  \begin{aligned}
    g_{\mu\nu}
    & =
    \hat{g}_{\mu\nu}-\hat{g}_{\underline{z}\mu}
    \hat{g}_{\underline{z}\nu}/\hat{g}_{\underline{z}\underline{z}}\, ,
    \\
    & \\
    A_{\mu}
    & =
    \hat{g}_{\mu \underline{z}}/\hat{g}_{\underline{z}\underline{z}}\, ,
    \\
    & \\
    k
    & =
    |\hat{g}_{\underline{z}\underline{z}}|^{1/2}\, ,
    \\
    & \\
    B^{(1)}{}_{\mu\nu}
    & =
    \hat{B}_{\mu\nu}
    +\hat{g}_{\underline{z}[\mu}
    \left[
         \hat{B}_{|\nu]\, \underline{z}}
      +\frac{\alpha'}{4}\left(\hat{A}^{A}{}_{|\nu]}\hat{A}_{A\, \underline{z}}
      +\hat{\Omega}^{(0)}_{(-)\, |\nu]}{}^{\hat{a}}{}_{\hat{b}}
      \hat{\Omega}^{(0)}_{(-)\, \underline{z}}{}^{\hat{b}}{}_{\hat{a}}
      \right)\right]
    /\hat{g}_{\underline{z}\underline{z}}\, ,
    \\
    & \\
    B^{(1)}{}_{\mu}
    & =
     \hat{B}_{\mu\, \underline{z}}
      -\frac{\alpha'}{4}\left[\hat{A}^{A}{}_{\mu}\hat{A}_{A\, \underline{z}}
      +\hat{\Omega}^{(0)}_{(-)\, \mu}{}^{\hat{a}}{}_{\hat{b}}
      \hat{\Omega}^{(0)}_{(-)\, \underline{z}}{}^{\hat{b}}{}_{\hat{a}}
    \right.
    \\
    & \\
    &
    \left.
      -\hat{g}_{\mu \underline{z}}\left(\hat{A}^{A}{}_{\underline{z}}\hat{A}_{A\, \underline{z}}
      +\hat{\Omega}^{(0)}_{(-)\, \underline{z}}{}^{\hat{a}}{}_{\hat{b}}
      \hat{\Omega}^{(0)}_{(-)\, \underline{z}}{}^{\hat{b}}{}_{\hat{a}}\right)
  /\hat{g}_{\underline{z}\underline{z}}\right]\, ,
    \\
    & \\
    \phi
    & =
    \hat{\phi}-\tfrac{1}{4}\log{(-\hat{g}_{\underline{z}\underline{z}})}\, ,
    \\
    & \\
    A^{A}{}_{\mu}
    & =
    \hat{A}^{A}{}_{\mu} -\hat{A}^{A}{}_{\underline{z}}\hat{g}_{\mu\underline{z}}/\hat{g}_{\underline{z}\underline{z}}\, ,
    \\
    & \\
    \varphi^{A}
    & =
    \hat{A}^{A}{}_{\underline{z}}/(-\hat{g}_{\underline{z}\underline{z}})^{1/2}\, .
  \end{aligned}
\end{equation}

\end{document}